\documentclass{aa}

\usepackage{graphicx}
\usepackage{natbib}
\usepackage{txfonts}
\usepackage[colorlinks=true,allcolors=blue]{hyperref}

\newcommand{\bz}{$\langle B_\mathrm{z} \rangle$}
\newcommand{\bm}{$\langle B \rangle$}
\newcommand{\mbz}{\langle B_\mathrm{z} \rangle}
\newcommand{\mbbz}{\langle B^2 B_\mathrm{z} \rangle}
\newcommand{\mbsq}{\langle B^2 \rangle_\mathrm{z}}
\newcommand{\teff}{$T_{\rm eff}$}

\newcommand{\kms}{km\,s$^{-1}$}

\newcommand{\equ}{$\gamma$\,Equ}
\newcommand{\aql}{10\,Aql}
\newcommand{\leo}{AD\,Leo}
\newcommand{\lac}{EV\,Lac}

\newcommand{\orcidlink}[1]{\protect\href{https://orcid.org/#1}{\protect\includegraphics[width=8pt]{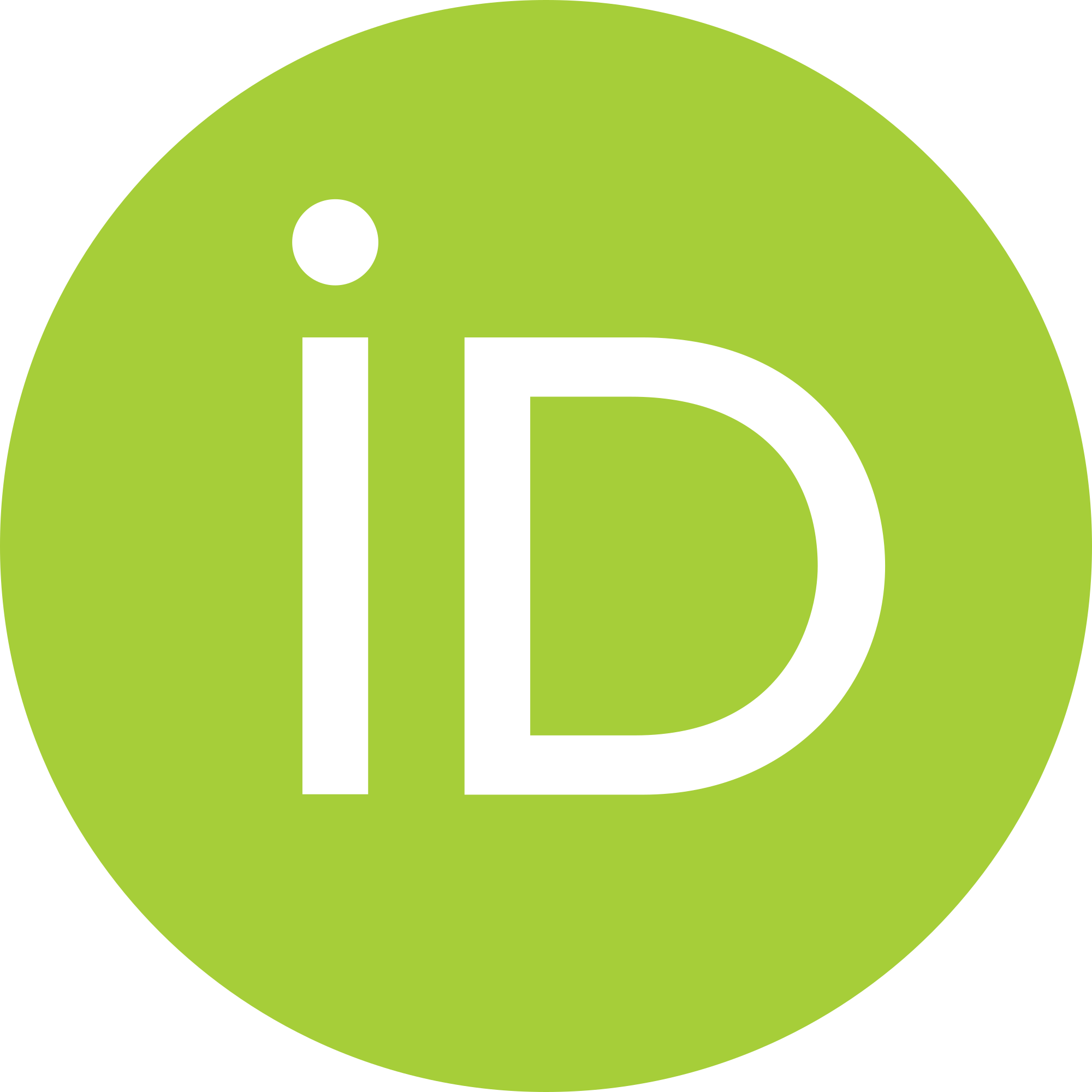}}}

\newcommand{\figps}[1]{\resizebox{\hsize}{!}{\rotatebox{0}{\includegraphics{#1}}}}

\newcommand{\fifps}[2]{\centering\resizebox{#1}{!}{\includegraphics{#2}}}

\newcommand{\beq}{\begin{equation}}
\newcommand{\eeq}{\end{equation}}

\begin{document}

\title{Unlocking new information in stellar circular polarisation with second-order least-squares deconvolution}

\author{%
Oleg Kochukhov\corrauth{oleg.kochukhov@physics.uu.se}\orcidlink{0000-0003-3061-4591}
}

\institute{
Department of Physics and Astronomy, Uppsala University, Box 524, SE-75120 Uppsala, Sweden
}

\date{Received 10 August 2026 / Accepted 2 October 2026}

\titlerunning{Second-order LSD}
\authorrunning{Oleg Kochukhov}

\abstract
{
Least-squares deconvolution (LSD) is the standard technique for measuring stellar
magnetic fields from circular polarisation spectra. It rests on the weak-field
approximation and yields the longitudinal field alone. The field modulus is
accessible only through a methodologically distinct analysis of intensity spectra, limited to strongly magnetic stars or slow
rotators.
}
{
We aim to extend LSD beyond the weak-field limit and interpret departures from a linear
field response in terms of the field modulus.
}
{
The Stokes $V$ absorption coefficient is expanded in odd wavelength derivatives
of the intensity profile, with coefficients given by the Zeeman pattern moments.
Truncation after two terms gives every line a second weight, formed from the
third moment of its splitting pattern, and two mean Stokes $V$ profiles follow from
multiprofile LSD. This description was tested with polarised spectrum
synthesis and applied to archival spectra of two Ap stars and two active M
dwarfs.
}
{
The second mean profile is proportional to the second derivative of the first,
with a coefficient measuring the mean square field modulus weighted by the
longitudinal flux. This observable separates field strength from the area
it covers.
Simulations at realistic signal-to-noise ratios show that the second component is
reliably detected from about 500 G and constrains the field modulus up to several kG.
Applying this methodology to the Ap stars, we found the modulus consistent with their known
field strengths, whereas for the M dwarfs it was close to the
Zeeman-broadening measurements and far above the large-scale fields from Zeeman-Doppler
imaging.
}
{
Second-order LSD turns the breakdown of the weak-field approximation from a source
of error into a new magnetic observable, revealing for the first time the field
strength in the regions emitting circular polarisation. It applies to any star
with a detectable Stokes $V$ signature and bridges the gap between polarimetry and Zeeman broadening,
most promisingly for cool active stars in the infrared.
}


\maketitle
\nolinenumbers

\section{Introduction}
\label{sec:intro}

High-resolution circular polarisation spectroscopy is the principal source of
information on the magnetic fields of non-degenerate stars. The longitudinal
Zeeman effect renders the circular polarisation profiles of magnetically sensitive lines antisymmetric
about their centres, with a Stokes $V$ amplitude proportional to the
line-of-sight component of the field. Unlike the Zeeman broadening of the
intensity profiles, Stokes $V$ remains measurable for fields far too weak to
alter the width of a line
\citep[e.g.][]{landstreet:1992,donati:2009,reiners:2012}. The Zeeman effect was
first detected in a stellar polarisation spectrum by \citet{babcock:1947}. The diagnostic has
since been applied across most of the Hertzsprung-Russell diagram, from the
strong fossil fields of Ap/Bp and massive OB stars
\citep[e.g.][]{wade:2000,wade:2016} to the few-gauss large-scale fields
of solar analogues \citep[e.g.][]{petit:2008a,metcalfe:2024,chiti:2025} and the complex topologies of fully convective M dwarfs
\citep[e.g.][]{donati:2006,morin:2010,kochukhov:2017c}.

A major fraction of this work is carried out on mean line profiles rather than on
individual lines. Least-squares deconvolution \citep[LSD,][]{donati:1997,kochukhov:2010a}
represents the observed polarisation spectrum as a superposition of identical
profiles, one for each line of a mask, scaled by weights formed from the central
depth, wavelength, and effective Land\'e factor of the transition. The mean
profile recovered in this way combines the signal of thousands of lines,
improving the noise level by a factor of order the square root of their number.
Detection thresholds thereby reach a few parts in $10^5$ of the continuum.
This gain is what makes circular polarisation studies of cool stars practicable,
since the Stokes $V$ signature of any single line typically lies well below the
noise of an individual exposure. LSD profiles have accordingly become the
standard data product of stellar spectropolarimetry, serving for detection
assessment, for longitudinal field measurements, and as input data for
the reconstruction of the surface field geometry with
Zeeman-Doppler imaging \citep[ZDI,][]{kochukhov:2016}.
Multiline alternatives based on principal component analysis
\citep{martinez-gonzalez:2008,carroll:2012}, Gaussian processes \citep{asensio-ramos:2015},
or nonlinear deconvolution \citep{sennhauser:2009a}
have also been developed, but none of them has displaced LSD in practice.

Two core approximations underlie the LSD technique. The first is the weak-field
limit, in which Stokes $V$ is proportional to the longitudinal field, to the
effective Land\'e factor, and to the wavelength derivative of Stokes $I$. The
second is self-similarity, with every line of the mask assumed to contribute the
same profile shape and to differ only by its scalar weight. Both hold while the
Zeeman splitting remains small compared with the intrinsic line width. However,
for optical lines of typical Land\'e factor the splitting reaches the Doppler
width already at about a kilogauss. This value lies well inside the range of local
field strengths measured in active FGK stars
\citep[e.g.][]{kochukhov:2020,kochukhov:2023}, and is exceeded routinely
in magnetic early-type stars \citep[e.g.][]{mathys:1997b,mathys:2017}
and in active M dwarfs \citep[e.g.][]{kochukhov:2021,reiners:2022}. The resulting distortion of the mean profile is a
recognised limitation on the quantitative interpretation of LSD Stokes $V$
\citep{kochukhov:2010a,kochukhov:2014,kochukhov:2026}. 

Two quantities therefore remain out of reach of standard first-order LSD. The first is
the field modulus, since circular polarisation constrains only the product of the
local field strength and the fraction of the surface that carries it. The modulus
has to be sought instead
in the Zeeman splitting measurements of early-type stars \citep[e.g.][]{mathys:1997b}
and in the Zeeman broadening or intensification of the
intensity profiles of late-type stars \citep[e.g.][]{robinson:1980,johns-krull:1996,kochukhov:2020}, both of
which require either strong fields or slow rotation. The second is the magnetic energy
stored in unresolved structures of mixed polarity, whose circular polarisation
signatures cancel in any disk-integrated observable, so that ZDI maps recover only
a small fraction of the true surface field energy
\citep[e.g.][]{reiners:2009,see:2019,kochukhov:2021}.

Departures from the linear scaling of the Stokes $V$ profile are not only a source of error but also a source of
information. For the solar case, \citet{stenflo:1973} showed that the ratio of the
Stokes $V$ amplitudes of two lines of the same multiplet with different Land\'e
factors separates the intrinsic field strength from the fraction of the resolution
element that the field occupies. It does so precisely because the response of
Stokes $V$ to the field ceases to be linear. In this study we develop the
multiline analogue of that idea, implementing it as a natural second-order extension of the LSD methodology.
This extension gives access to the strength of the field itself, and not
merely to the flux that it carries, from circular polarisation alone. We show that such a
measurement can be carried out for the moderate fields
that leave no detectable trace in the intensity spectrum, and that it responds to the
small-scale magnetic energy that standard first-order LSD cannot represent.
The technique we propose calls for no new
observations and no new instruments, so that the two families of magnetic
diagnostics applied so far to different stars by different means may be brought
together on data that already exist.

The paper is organised as follows. Sect.~\ref{sec:method} sets out the
theoretical basis, deriving the second-order description of the multiline Stokes
$V$ profile from the moments of the Zeeman pattern through the
expansion of the absorption coefficient to the integration over the visible
stellar hemisphere. Sect.~\ref{sec:diag} discusses what the
additional mean profile contributes to the detection of magnetic fields, 
to disk-averaged magnetic observables, and to ZDI mapping,
together with the practical requirements of the method. 
Sect.~\ref{sec:val} tests the description against
polarised spectrum synthesis, first for unblended local profiles
and then for blended disk-integrated spectra. The main results are
summarised and their implications discussed in Sect.~\ref{sec:concl}.
Preliminary application of the methodology developed here to cool Ap stars and active M dwarfs is presented in Appendix~\ref{sec:app}.

\section{Theoretical basis}
\label{sec:method}

\subsection{Least-squares deconvolution and the weak-field limit}
\label{sec:lsd1}

The least-squares deconvolution technique 
represents a polarisation spectrum $Y(\varv)$ as a superposition of identical profiles
$Z(\varv)$, each scaled by a line-dependent weight $w_i$ and shifted to the
position $\varv_i$ of the corresponding line in the mask,
\begin{equation}
Y(\varv) = \sum_i w_i\, Z(\varv-\varv_i)\,,
\label{eq:lsd}
\end{equation}
or $Y = \mathbf{M}\cdot Z$ in matrix form. The line pattern matrix
$\mathbf{M}$, given by Eq. (17) of \citet{kochukhov:2010a}, 
is constructed from the line list, and the common profile $Z$ is
recovered by linear least squares.

For circular polarisation this description rests on the weak-field
approximation. If the Zeeman splitting of the line is much smaller than the intrinsic
width of the absorption profile, its Stokes $V$ spectrum is
proportional to the wavelength derivative of Stokes $I$
\citep[e.g.][]{polarization:2004},
\begin{equation}
V(\lambda) = -\Delta\lambda_{\rm B}\,\bar{g}\cos\theta\,
\frac{\partial I}{\partial\lambda}\,,
\label{eq:weakfield}
\end{equation}
where $\bar{g}$ is the effective Land\'e factor and $\theta$ is the angle between
the magnetic field vector and the line of sight. The quantity
\begin{equation}
\Delta\lambda_{\rm B} = \frac{\lambda_0^2 e_0 B}{4\pi m_{\rm e}c^2}
\equiv k\lambda_0^2 B \,,
\qquad k = 4.6686\times10^{-13}\,\mathrm{\AA^{-1}\,G^{-1}}
\label{eq:dlb}
\end{equation}
is the Lorentz unit for a line of central wavelength $\lambda_0$, with $B$
the field modulus, $e_0$ the absolute value of the electron charge, and
$m_{\rm e}$ the electron mass.
Since the residual line profile, $1-I/I_{\rm c}$, scales with the central
line depth $d$, and
$\partial/\partial\lambda \propto \lambda_0^{-1}$ when the profile is expressed
in velocity units, Eq.~(\ref{eq:weakfield}) yields the familiar Stokes $V$
weight $w = d\lambda_0\bar{g}$.

Equation~(\ref{eq:weakfield}) is the leading term of an expansion in the
dimensionless Zeeman splitting
\begin{equation}
\zeta = \frac{\Delta\lambda_{\rm B}}{\Delta\lambda_{\rm D}}\,,
\label{eq:zeta}
\end{equation}
where $\Delta\lambda_{\rm D}$ is the intrinsic Doppler width of the
absorption profile, set by the thermal and microturbulent velocities alone, with
rotational, macroturbulent, and instrumental broadening excluded.
The validity of the weak-field approximation is
conventionally stated as $\bar{g}\,\zeta\ll1$. This expansion does not terminate
there, and its next non-vanishing term can be accommodated
within the LSD formalism at the cost of a single additional common profile,
using the multiprofile capability already available in the \textsc{iLSD}\footnote{\url{https://github.com/okochukhov/iLSD}}
implementation of \citet{kochukhov:2010a}.

\subsection{Moments of the Zeeman pattern}
\label{sec:moments}

A spectral line connecting a lower level with the total angular momentum quantum number $J_\ell$ and Land\'e factor $g_\ell$ to an upper level
$J_u$ and $g_u$ splits into components displaced from the line centre by
\begin{equation}
\Delta\lambda_{M_\ell M_u} = \left(g_\ell M_\ell - g_u M_u\right)
\Delta\lambda_{\rm B}
\label{eq:shift}
\end{equation}
with relative strengths 
\begin{equation}
S_q(M_\ell,M_u) = 3\begin{pmatrix} J_u & J_\ell & 1\\
-M_u & M_\ell & -q\end{pmatrix}^{2}\,,
\label{eq:strength}
\end{equation}
where the array denotes a Wigner $3j$ symbol, and $q = M_\ell-M_u = -1,0,+1$
labels the $\sigma_{\rm b}$, $\pi$, and $\sigma_{\rm r}$ groups. The factor 3
normalises each group to unity. Following \citet{polarization:2004}, the $n$-th moment of the
$q$-group is written as
\begin{equation}
\sum_{M_\ell M_u} S_q(M_\ell,M_u)\left[\Delta\lambda_{M_\ell M_u}\right]^n
= \Delta\lambda_{\rm B}^n\, G^{(n)}_q\,,
\label{eq:moments}
\end{equation}
so that $G^{(n)}_q$ is simply that moment in units of
$\Delta\lambda_{\rm B}$. These quantities obey
\begin{equation}
G^{(n)}_{-q} = (-1)^n\, G^{(n)}_q\,,
\label{eq:sym}
\end{equation}
which follows from the symmetry of the pattern about the line centre.

The first
moment of the $\sigma_{\rm r}$ group is the effective Land\'e factor,
\begin{equation}
G^{(1)}_{1} = \bar{g} = \frac{1}{2}\left(g_\ell+g_u\right)
+ \frac{1}{4}\left(g_\ell-g_u\right)\delta\,,
\label{eq:gbar}
\end{equation}
while the third moment, required below, is
\begin{equation}
\begin{split}
G^{(3)}_{1} = \frac{1}{80} \Big[& \left(3\delta^2-7\right)\delta\,g_{\rm d}^3
+ 6 \left(\delta^2+2s-3\right)g_{\rm d}^2 g_{\rm s} \\
&+ 15\,\delta\,g_{\rm d}g_{\rm s}^2 + 10\,g_{\rm s}^3\Big]\,,
\end{split}
\label{eq:g3}
\end{equation}
with $s = J_\ell(J_\ell+1)+J_u(J_u+1)$, $\delta = J_\ell(J_\ell+1)-J_u(J_u+1)$,
$g_{\rm s} = g_\ell+g_u$, and $g_{\rm d} = g_\ell-g_u$. Whenever the
$\sigma_{\rm r}$ group reduces to a single wavelength, as it does for a Zeeman
triplet, $G^{(n)}_1 = \bar{g}^{\,n}$ for every $n$. The departure of
$G^{(3)}_1$ from $\bar{g}^{\,3}$ thus measures the anomalous character of
the splitting pattern, and constitutes atomic information that $\bar{g}$ alone
does not determine. Both $\bar{g}$ and $G^{(3)}_1$ follow from
the four quantities $J_\ell$, $J_u$, $g_\ell$, and $g_u$, available in atomic
line lists such as those provided by the VALD database
\citep{ryabchikova:2015}. Where a Land\'e factor is not tabulated it may be
estimated from the term designation of the level under the coupling scheme VALD
assigns to it (LS, JJ, or JK).

\subsection{Expansion of the Stokes $V$ absorption coefficient}
\label{sec:etav}

Following the perturbative scheme of \citet{polarization:2004}, the
expansion in field strength is organised by replacing $B$ with $\alpha B$, at
fixed field direction, and developing every quantity in powers of $\alpha$. This
parameter carries no physical content, since $\alpha=0$ is the
field-free atmosphere and $\alpha=1$ the actual one, so that expanding about
$\alpha=0$ is an expansion about the unmagnetised case, in which the coefficient
of $\alpha^n$ collects all contributions proportional to the $n$-th power of the
field. The quantity expanded here is the absorption coefficient of each Zeeman
group. Because every Zeeman component is a
displaced copy of the same zero-field absorption profile
$\eta(\lambda)$, with displacements proportional to $\alpha$, the profile
of the $q$-group is
\begin{equation}
\eta_q(\alpha) = \sum_{n=0}^{\infty} \frac{(-\alpha)^n}{n!}\,
G^{(n)}_q\, \Delta\lambda_{\rm B}^n\,
\frac{\partial^n\eta}{\partial\lambda^n}\,,
\label{eq:etaq}
\end{equation}
where Eq.~(\ref{eq:moments}) has been used to carry out the sum over magnetic
sublevels. Inserting this into the Stokes $V$ element of the absorption matrix,
$h_V = \frac{1}{2}\left(\eta_{+1}-\eta_{-1}\right)\cos\theta$, the even orders
cancel according to Eq.~(\ref{eq:sym}) and the odd ones double, leaving
\begin{equation}
h_V = -\cos\theta \sum_{n\,\mathrm{odd}}
\frac{\alpha^n \Delta\lambda_{\rm B}^n}{n!}\,G^{(n)}_1\,
\frac{\partial^n\eta}{\partial\lambda^n}\,.
\label{eq:hv}
\end{equation}

Equation~(\ref{eq:hv}) is exact: it is the Taylor expansion in the component
displacements of Eq.~(\ref{eq:shift}), which translate $\eta$ without altering
its shape. Two of its properties
are essential for what follows. The expansion coefficients
$\Delta\lambda_{\rm B}^n G^{(n)}_1/n!$ are fixed entirely by the atomic
parameters of the transition and by the field strength, the shape of the
profile entering only through the derivatives of $\eta$. The Doppler width
is thus absent from the coefficients, and enters only through
$\partial^n\eta/\partial\lambda^n \propto \Delta\lambda_{\rm D}^{-n}$.
Measuring the derivatives in units of the profile width therefore recovers the expansion
in $\zeta$ of Eq.~(\ref{eq:zeta}), with coefficients $G^{(n)}_1/n!$ and shape
factors of order unity. Rotational, macroturbulent, and instrumental
broadening modify these derivatives but, since differentiation commutes with
convolution, leave the coefficients unchanged.

Retaining the first two non-vanishing terms and setting $\alpha=1$,
\begin{equation}
h_V \simeq -\Delta\lambda_{\rm B}\,\bar{g}\cos\theta\,
\frac{\partial\eta}{\partial\lambda}
- \frac{1}{6}\Delta\lambda_{\rm B}^3\,G^{(3)}_1\cos\theta\,
\frac{\partial^3\eta}{\partial\lambda^3}\,.
\label{eq:hv13}
\end{equation}
Successive terms of Eq.~(\ref{eq:hv}) decrease by a factor of order $\left(\bar{g}\zeta\right)^2$,
so the second term of Eq.~(\ref{eq:hv13}) is the leading correction to the
weak-field description given by the first term.

\subsection{Emergent Stokes $V$ profile}
\label{sec:transfer}

Substituting the expansions of the propagation matrix into the polarised radiative
transfer equation and collecting equal powers of $\alpha$ yields a hierarchy of
equations for the expansion coefficients of the Stokes parameters,
$V = V_1\alpha + V_3\alpha^3 + \dots$ and $I = I_0 + I_2\alpha^2 + \dots$
\citep{polarization:2004}. The two orders of interest are
\begin{equation}
\frac{{\rm d}V_1}{{\rm d}\tau_{\rm c}} = \left(1+\kappa_{\rm L}\eta\right)V_1
- \Delta\lambda_{\rm B}\,\bar{g}\cos\theta\,\kappa_{\rm L}\,\eta'
\left(I_0-S_{\rm L}\right)
\label{eq:v1}
\end{equation}
and
\begin{equation}
\begin{split}
\frac{{\rm d}V_3}{{\rm d}\tau_{\rm c}} =\;&
\left(1+\kappa_{\rm L}\eta\right)V_3
- \frac{1}{6}\Delta\lambda_{\rm B}^3\,G^{(3)}_1\cos\theta\,
\kappa_{\rm L}\,\eta'''\left(I_0-S_{\rm L}\right) \\
&- \Delta\lambda_{\rm B}\,\bar{g}\cos\theta\,\kappa_{\rm L}\,\eta'\,I_2 
+ \frac{1}{4}\Delta\lambda_{\rm B}^2\,\Gamma\,\kappa_{\rm L}\,\eta''\,V_1 \,.
\end{split}
\label{eq:v3}
\end{equation}
The intensity correction $I_2$ entering the second source term obeys in turn
\begin{equation}
\begin{split}
\frac{{\rm d}I_2}{{\rm d}\tau_{\rm c}} =\;&
\left(1+\kappa_{\rm L}\eta\right)I_2
+ \frac{1}{4}\Delta\lambda_{\rm B}^2\,\Gamma\,\kappa_{\rm L}\,\eta''
\left(I_0-S_{\rm L}\right) \\
&- \Delta\lambda_{\rm B}\,\bar{g}\cos\theta\,\kappa_{\rm L}\,\eta'\,V_1 \,,
\end{split}
\label{eq:i2}
\end{equation}
where $\tau_{\rm c}$ is the continuum optical depth, $\kappa_{\rm L}$ the ratio
of line to continuum opacity, $S_{\rm L}$ the line source function,
\begin{equation}
\Gamma = G^{(2)}_0\sin^2\theta + G^{(2)}_1(1+\cos^2\theta),
\label{eq:Gamma}
\end{equation}
and primes denote derivatives with respect to wavelength.
The magneto-optical terms of the propagation matrix do not appear, entering
Stokes $V$ only at order $\alpha^4$.

These three source terms are all of the same order in the
magnetic field. Since $V_1 \propto \zeta$ and $I_2 \propto \zeta^2$, the
explicit powers of $\Delta\lambda_{\rm B}$ multiplying them bring the second and
third terms to order $\zeta^3$, the same as the first. They
are distinguished instead by their dependence on the line opacity. Consider the
weak-line limit $\kappa_{\rm L}\eta\ll1$, in which the attenuation along the ray
becomes wavelength independent and any source of the form $\kappa_{\rm
L}f(\lambda)g(\tau_{\rm c})$ factorises. Writing
\begin{equation}
\mathcal{K} = -\int_0^{\infty}\kappa_{\rm L}\left(I_0-S_{\rm L}\right)
{\rm e}^{-\tau_{\rm c}}\,{\rm d}\tau_{\rm c} \;\propto\; \kappa_{\rm L}\,,
\label{eq:kfac}
\end{equation}
the emergent values of the lower-order coefficients are
\begin{equation}
V_1 \simeq -\Delta\lambda_{\rm B}\,\bar{g}\cos\theta\;\mathcal{K}\,\eta'\,,
\qquad
I_2 \simeq \frac{1}{4}\Delta\lambda_{\rm B}^2\,\Gamma\;\mathcal{K}\,\eta''\,,
\label{eq:v1i2}
\end{equation}
both of order $\kappa_{\rm L}$. The last two source terms of Eq.~(\ref{eq:v3}) are thus
of order $\kappa_{\rm L}^2$, whereas the first is of order $\kappa_{\rm L}$.
To leading order in the line opacity the emergent circular polarisation therefore
reproduces the absorption coefficient,
\begin{equation}
V = V_1 + V_3 = \mathcal{K}\,h_V\,,
\label{eq:vhv}
\end{equation}
and inherits the two-term structure of Eq.~(\ref{eq:hv13}).

The weak-line assumption is required only for the third-order
term. At first order the exact result of \citet{polarization:2004},
Eq.~(\ref{eq:weakfield}), holds for lines of arbitrary strength provided the
line-of-sight field does not vary with depth, so the
first-order term coincides with standard LSD and remains untouched by the
weak-line assumption.

\subsection{The second-order expansion in velocity units}
\label{sec:lsd2}

Equation~(\ref{eq:hv13}) is conveniently expressed in velocity units,
$\varv = c(\lambda-\lambda_0)/\lambda_0$, so that
$\partial/\partial\lambda = (c/\lambda_0)\,\partial/\partial\varv$ and
$\Delta\lambda_{\rm B}\,c/\lambda_0 = kc\lambda_0B$,
\begin{equation}
h_V(\varv) = -kcB\cos\theta\;\lambda_0\left[\bar{g}\,
\frac{\partial\eta}{\partial\varv}
+ \frac{1}{6}\left(kcB\right)^2\lambda_0^2\,G^{(3)}_1\,
\frac{\partial^3\eta}{\partial\varv^3}\right].
\label{eq:hvv}
\end{equation}
The coefficients of the two terms in the square brackets differ by the
product of two factors. The first of these,
$\lambda_0^2\,G^{(3)}_1/\bar{g}$, depends only on the line. The second,
\begin{equation}
\gamma = \frac{1}{6}\left(kcB\right)^2 ,
\label{eq:gamma}
\end{equation}
depends only on the magnetic field, or numerically
$\gamma = 3.2649\times10^{-15}\,B^2$ km$^2$\,s$^{-2}$\,\AA$^{-2}$ for $B$
expressed in G. This separation
is precisely what the LSD formalism requires of a line weight.

These two terms in Eq.~(\ref{eq:hvv}) are, moreover, the first and the third
derivative of one and the same function, $\partial^3\eta/\partial\varv^3$ being
the second velocity derivative of $\partial\eta/\partial\varv$. The second term is
obtained from the first by differentiating it twice with respect to velocity
and multiplying the result by the line-dependent factor $\lambda_0^2\,G^{(3)}_1/\bar{g}$
and by $\gamma$, which is proportional to the square of the field strength.
The two contributions to $h_V$ are thus not independent functions of velocity.

\subsection{Disk-integrated mean profiles}
\label{sec:disk}

Let $\Omega$ denote a point of the visible stellar hemisphere, $B(\Omega)$ the local field
modulus, $B_\mathrm{z}(\Omega) = B(\Omega)\cos\theta(\Omega)$ its line-of-sight component,
$b(\Omega)$ the local line-depression flux (the product of the continuum and
line-strength centre-to-limb variations, both common to all lines), normalised to
$\int b\,{\rm d}\Omega = 1$, $I_{\rm c}$ the disk-integrated continuum, and
$\varv_{\Omega}$ the Doppler shift produced by rotation.

Because $\mathcal{K}$ is independent of wavelength in the weak-line limit,
the derivatives in Eq.~(\ref{eq:hvv}) act on the combination $\mathcal{K}\eta$,
which is the wavelength-dependent part of the local Stokes $I$ profile,
its depression below the local continuum, negative in an absorption line because $\mathcal{K}<0$. Two assumptions are now
required, and they are precisely those on which standard
LSD already rests \citep{kochukhov:2010a}. The line
depression $-\mathcal{K}\eta$
is described by a common shape $\mathcal{I}(\varv)$ scaled by the central
line depth $d_i$, and the centre-to-limb variation of the profiles is
the same for all
lines. Both terms of Eq.~(\ref{eq:hvv}) carry a single factor $\cos\theta$
while carrying $B$ and $B^3$ respectively, so the local first- and third-order
contributions are proportional to $B_\mathrm{z}$ and to $B^2B_\mathrm{z}$.
Integrating over the visible hemisphere, the contribution of the $i$-th line to
the observed circular polarisation spectrum is
\begin{equation}
\frac{V_i(\varv)}{I_{\rm c}} = w^{(1)}_i\,Z_1(\varv)
+ w^{(3)}_i\,Z_3(\varv)\,,
\label{eq:model}
\end{equation}
where all the line-specific information is carried by the two sets of weights
\begin{equation}
w^{(1)}_i = d_i\,\lambda_{0,i}\,\bar{g}_i\,,
\qquad
w^{(3)}_i = d_i\,\lambda_{0,i}^3\,G^{(3)}_{1,i}\,,
\label{eq:weights}
\end{equation}
and the two mean profiles common to all lines are
\begin{align}
Z_1(\varv) &= kc\int b(\Omega)\,B_\mathrm{z}(\Omega)\,
\frac{\partial\mathcal{I}}{\partial\varv}(\varv-\varv_{\Omega})\,{\rm d}\Omega\,,
\label{eq:z1}\\
Z_3(\varv) &= \frac{1}{6}(kc)^3\int b(\Omega)\,B^2(\Omega)B_\mathrm{z}(\Omega)\,
\frac{\partial^3\mathcal{I}}{\partial\varv^3}(\varv-\varv_{\Omega})\,{\rm d}\Omega\,.
\label{eq:z3}
\end{align}

Equations~(\ref{eq:model})--(\ref{eq:z3}) are the central result of this
section. The two mean profiles are built from the same local residual
profile by a different number of differentiations, and are weighted over the stellar
surface by $B_\mathrm{z}$ in one case and by $B^2B_\mathrm{z}$ in the other.
This difference in
weighting is what carries the additional magnetic information, and everything in
Sect.~\ref{sec:diag} follows from it. We denote the corresponding
disk averages by
\begin{equation}
\mbz = \int b\,B_\mathrm{z}\,{\rm d}\Omega\,,
\qquad
\mbbz = \int b\,B^2B_\mathrm{z}\,{\rm d}\Omega\,.
\label{eq:diskav}
\end{equation}

The profiles $Z_1$ and $Z_3$ are in general two independent functions of
velocity. The
scaling $\gamma$ can be taken outside the integrals in Eqs.~(\ref{eq:z1}) and
(\ref{eq:z3}) in three circumstances, leading to
\begin{equation}
Z_3(\varv) = \gamma\,\frac{\partial^2 Z_1}{\partial\varv^2}\,.
\label{eq:tie}
\end{equation}
The first situation in which this expression is valid is a field of constant modulus, for which $\gamma$ is that of
Eq.~(\ref{eq:gamma}). The second is a surface every point of which carries the
same Doppler shift. Here $\partial^3\mathcal{I}/\partial\varv^3$ factors out of
both $Z_3$ and $\partial^2Z_1/\partial\varv^2$, and $\gamma = \frac{1}{6}(kc)^2\,\mbbz/\mbz$, whatever the
distribution of field strengths. 
The third is a modulus constant over the regions where $B_\mathrm{z}$ does
not vanish, whatever their extent and distribution over the stellar surface,
$\gamma$ again being that of Eq.~(\ref{eq:gamma}).
A departure from Eq.~(\ref{eq:tie}) therefore requires a modulus that varies
over those regions and a spread of Doppler shifts over the visible hemisphere.

Equation~(\ref{eq:model}) has exactly the form of the multiprofile
generalisation of LSD \citep{kochukhov:2010a}: $Z_1$ and $Z_3$ are recovered
simultaneously by linear least squares from the composite line pattern matrix
$\mathbf{M} = [\mathbf{M}_1\;\mathbf{M}_3]$ built with the two weight sets of
Eq.~(\ref{eq:weights}). The first of these, $w^{(1)}$, is identical to the
weight used in standard LSD, so the conventional mean profile is recovered
whenever the second component is omitted. Since $Z_3$ is weighted over the
surface by $B^2B_\mathrm{z}$ whereas $Z_1$ is weighted by $B_\mathrm{z}$, the
second contribution relative to the first scales as the square of the field strength, so that the
two-component model tends to standard LSD in the weak-field limit.

\section{Second-order magnetic field diagnostics}
\label{sec:diag}

Standard LSD supports a hierarchy of magnetic measurements: the detection of a
circular polarisation signature, the determination of the mean longitudinal
field from the first moment of the mean profile, and the modelling of time
series of mean profiles with ZDI. The availability
of a second mean profile modifies each of them.

\subsection{Detection of the second component}
\label{sec:detect}

The profile $Z_3$ changes sign under a
reversal of the local field, exactly as $Z_1$ does, because the surface weightings
$B_\mathrm{z}$ and $B^2B_\mathrm{z}$ of Eqs.~(\ref{eq:z1}) and (\ref{eq:z3}) are
both odd. The contributions of two regions of equal area and opposite polarity, seen at the same
limb angle and carrying the same Doppler shift, therefore cancel in $Z_1$, and
they cancel in $Z_3$ as well provided that the two regions also carry the same field
modulus. The second component thus gives no access to the unsigned magnetic flux
and will not by itself reveal a field that standard LSD misses for this reason. However, where
the two polarities carry different distributions of field strength their
contributions cancel in $Z_1$ and not in $Z_3$, so that a detection of the polarisation signature in the second
component without one in the first is physically plausible. Only mixed
polarity at the same field strength is invisible in both channels.

Its value at this level lies elsewhere. A significant $Z_3$ is a
model-independent indication that the weak-field approximation has broken down
for the star in question, and thus that ZDI reconstructions, and any
analysis relying on fitting the shape of the mean profile, require
revision. Standard LSD carries no internal warning of this.
Furthermore, the second component
is not free to take an arbitrary shape. It is built from the same local profile
as $Z_1$, so it occupies the same velocity interval, and both its area and its
first moment vanish (see Sect.~\ref{sec:moments2}), whereas for $Z_1$ only the area does. If the field modulus is
in addition nearly uniform over the regions where $B_\mathrm{z}$ does not vanish, or the rotational
broadening is small compared with the width of the local profile, $Z_3$ is tied
to $\partial^2Z_1/\partial\varv^2$ by the single number $\gamma$
(Eq.~\ref{eq:tie}). Their joint detection is then a considerably more stringent
test against instrumental artefacts than the detection of $Z_1$ alone.

A non-detection of $Z_3$ accompanied by a
detection of $Z_1$ is itself a measurement. Standard LSD cannot distinguish a
longitudinal field of, say, 10~G produced by a weak field covering the entire
surface from the same value produced by a kG field occupying 1\% of
it. The two configurations predict $Z_3$ amplitudes differing by four orders
of magnitude, so an upper limit on $Z_3$ translates into an upper limit on the
field modulus in the flux-carrying regions. This is likely the most useful
application of the second component in practice.

\subsection{Longitudinal field and mean square field modulus}
\label{sec:moments2}

The longitudinal field is measured from the first moment of the mean
profile, so the question is whether the second component interferes with
it. For $m \le n$, $m$ integrations by parts with vanishing
boundary terms give
\begin{equation}
\int \varv^m\,\frac{\partial^n f}{\partial\varv^n}\,{\rm d}\varv =
\begin{cases}
0\,, & m < n\,,\\[2pt]
(-1)^n\,n!\int f\,{\rm d}\varv\,, & m = n\,.
\end{cases}
\label{eq:momzero}
\end{equation}
The first moment of Stokes $V$, the case $m = 1$, therefore samples the $n=1$ term of the
expansion (\ref{eq:hv}) alone. Taken over the full velocity range this
holds exactly, at all orders and for any
distribution of the field over the stellar surface. The second component
contributes nothing to the standard estimator of \bz\ and cannot
bias it. Whatever residual inaccuracy that estimator has must thus originate
outside the Zeeman expansion.

Inserting Eqs.~(\ref{eq:z1}) and (\ref{eq:z3}) into the first and the third
moment of the two profiles and interchanging the velocity and surface
integrations, each surface element enters with the moment of its own shifted
profile. Writing $\varv = (\varv-\varv_\Omega)+\varv_\Omega$ and expanding, every
term but the highest vanishes according to the first line of Eq.~(\ref{eq:momzero}). The
velocity integral therefore gives the same factor $-W_I$ for $Z_1$ and $-6W_I$
for $Z_3$ at every point of the disk, whatever its Doppler shift, the 6
cancelling the $1/6$ of Eq.~(\ref{eq:z3}). With the definitions of
Eq.~(\ref{eq:diskav}) and the weights of Eq.~(\ref{eq:weights}),
\begin{equation}
\begin{aligned}
\mbz &= -\frac{1}{kc\,W_I}\int \varv\,Z_1\,{\rm d}\varv \,,\\
\mbbz &= -\frac{1}{(kc)^3\,W_I}\int \varv^3 Z_3\,{\rm d}\varv \,,
\end{aligned}
\label{eq:moments13}
\end{equation}
where $W_I = \int\mathcal{I}\,{\rm d}\varv$ is the equivalent width
of the local profile. Owing to the earlier definition of
$b(\Omega)$, $W_I$ also equals the equivalent width of the
disk-integrated mean Stokes $I$ profile, directly accessible from the observed LSD intensity profile.
The first of these relations is the classical one underlying the standard LSD
longitudinal field estimator \citep[e.g.][]{kochukhov:2010a}. The factor $W_I$ is common to both, so the second calls
for no normalisation that the first does not already require. 

The ratio of the two disk averages defines a new observable,
\begin{equation}
\mbsq \equiv \frac{\mbbz}{\mbz} \,,
\label{eq:bsqz}
\end{equation}
which we refer to as the longitudinal-flux-weighted mean square field modulus.
Because $W_I$ enters the two relations of Eq.~(\ref{eq:moments13}) in the
same way, it cancels in their ratio, and $\mbsq$ can be formed without any Stokes
$I$ information, unlike \bz\ itself.
The interest of Eq.~(\ref{eq:bsqz}) lies in its behaviour for an inhomogeneous
surface. If a fraction $f$ of the stellar surface carries a field of modulus
$B_0$ and the remainder is field-free, the two averages of
Eq.~(\ref{eq:diskav}) differ only by a factor $B_0^2$, so that
\begin{equation}
\mbsq = B_0^2
\label{eq:filling}
\end{equation}
independently of $f$. Standard LSD constrains only the product $fB_0$, whereas the
second component removes this degeneracy, the two measurements together
determining the field strength and the filling factor separately. Equation
(\ref{eq:filling}) is the multiline analogue of the Stokes $V$ line-ratio
technique introduced for the solar case by \citet{stenflo:1973}.

Four limitations should be noted. Equation~(\ref{eq:bsqz}) becomes
ill-conditioned as $\mbz\rightarrow0$, so the ratio should not be formed at
individual rotational phases near a crossover, the modulations of \bz\ and
$\mbbz$ being better fitted jointly. The $\varv^3$ weighting in
Eq.~(\ref{eq:moments13}) amplifies noise in the profile wings far more strongly
than the first moment does, making that estimator sensitive to the integration
limits and to the continuum normalisation. However, where Eq.~(\ref{eq:tie}) holds,
that is for a modulus nearly uniform over the regions where $B_\mathrm{z}$ does not vanish or a rotational broadening small
compared with the width of the local profile, fitting $Z_3$ against
$\partial^2Z_1/\partial\varv^2$ over the entire profile is much better
conditioned. Where it does not, that fit returns
a mean of $B^2$ weighted by the overlap of the two profiles as well as by the
longitudinal flux, and it is Eq.~(\ref{eq:moments13}), noisy but exact, that
measures $\mbbz$. A single pair $(\mbz, \mbsq)$ does not determine the
inclination of the field and the distribution of its modulus separately, since an
elevation of $\mbsq$ above the value expected for a homogeneous longitudinal field
is produced by an inclined homogeneous field as readily as by an inhomogeneous
one. Separating the two requires the rotational modulation of both quantities or a
surface geometry from the ZDI inversion discussed in Sect.~\ref{sec:zdi}.
Finally, the strong-line corrections neglected in
Sect.~\ref{sec:transfer} carry angular factors that are not proportional to
$\cos\theta$, so for saturated lines Eq.~(\ref{eq:bsqz}) mixes $\mbbz$ with
$\langle B_\mathrm{z}^3\rangle$. These corrections are of order
$\kappa_{\rm L}^2$ and therefore negligible for weak lines, which is one of the
reasons to restrict the mask in central depth (Sect.~\ref{sec:practical}).

A single number need not be the end point. The functions
$\partial^2Z_1/\partial\varv^2$ and $Z_3$ are convolutions of the same kernel
$\partial^3\mathcal{I}/\partial\varv^3$ with two different surface weightings.
If the star rotates fast enough for the disk to be resolved in velocity, their
ratio in the deconvolved domain yields $\langle B^2\rangle$ as a function of
$\varv_{\Omega}$, and any departure of $Z_3$ from proportionality to
$\partial^2Z_1/\partial\varv^2$ is direct evidence of a distribution of field
strengths. Without that resolution Eq.~(\ref{eq:tie}) holds whatever the
distribution, and Eq.~(\ref{eq:bsqz}) is all that can be measured.

\subsection{Zeeman-Doppler imaging}
\label{sec:zdi}

The second-order circular polarisation LSD profile bears on ZDI in three ways.
The first is an improvement to the first-order profile itself: with a
single weight set the standard least-squares fit can
represent the $Z_3$ signal only by distributing it over $Z_1$. The mean
Stokes $V$ profile presented to the inversion is then not the true first-order one.
Its first moment is preserved, as Sect.~\ref{sec:moments2} shows, but its shape
and amplitude are not (Sect.~\ref{sec:accuracy}).
Recovering $Z_1$ together with $Z_3$ removes that distortion, and the gain
does not depend on $Z_3$ entering the inversion at all, since $Z_1$ from the
two-component solve is simply the better profile to invert.

Used as ZDI input data in its own right, $Z_3$ is a second information channel. Minimising
$\chi^2(Z_1)+\chi^2(Z_3)$ constrains one surface distribution through two
weightings, one linear and one cubic in the field. The forward problem changes
little, requiring the same local profile differentiated twice more and integrated with
weight $B^2B_\mathrm{z}$ in place of $B_\mathrm{z}$.
The gain is that the two channels respond differently to
the field strength inside a resolution element, which $Z_1$ alone cannot
separate from the area that carries it. This is the degeneracy that
Eq.~(\ref{eq:filling}) breaks for the disk-integrated quantities, and a
joint inversion should in principle break it for spatially resolved magnetic maps.

The third and most interesting consequence concerns the small-scale magnetic energy that
ZDI is known to miss. Writing the surface field as the sum of a large-scale
component $B_\mathrm{L}$ that ZDI resolves and a small-scale component
$B_\mathrm{S}$ whose polarities cancel gives
$\mbz \simeq \langle B_{\mathrm{L},\mathrm{z}}\rangle$ and
$\mbbz \simeq \langle B_\mathrm{S}^2 B_{\mathrm{L},\mathrm{z}}\rangle
+ \langle B_\mathrm{L}^2 B_{\mathrm{L},\mathrm{z}}\rangle$.
In this decomposition the term
$\langle B_\mathrm{S}^2B_{\mathrm{L},\mathrm{z}}\rangle$ does not
cancel. The small-scale field energy is even in $B_\mathrm{S}$ and
modulates the large-scale polarity pattern. Consequently, $\mbsq$ retains
sensitivity to the total $\langle B^2\rangle$, including the unresolved
component, and does so through circular polarisation, so that it remains
available for slowly rotating stars and moderate fields for which the Zeeman
broadening of Stokes $I$ is undetectable. Even before a joint inversion is
implemented, this suggests a simple consistency test. Predicting $Z_3$ from a
map reconstructed from $Z_1$ alone and comparing it with the observed $Z_3$
quantifies the small-scale energy that the map is missing. The limitation of
Sect.~\ref{sec:detect} nevertheless remains, since mixed polarities at fixed modulus
cancel in both channels, so that the method bridges ZDI and Zeeman broadening
rather than replacing either.

\subsection{Practical considerations}
\label{sec:practical}

In its basic form, this second-order multiline Stokes $V$ diagnostic
requires no modification of existing multiprofile LSD codes such as \textsc{iLSD}
\citep{kochukhov:2010a}.
The line mask is built as usual, with $\bar{g}$ and $G^{(3)}_1$ evaluated from
$(J_\ell,J_u,g_\ell,g_u)$ using Eqs.~(\ref{eq:gbar}) and (\ref{eq:g3}). Two
weight sets are then formed according to Eq.~(\ref{eq:weights}) and concatenated into
a composite mask, from which $Z_1$ and $Z_3$ are obtained by a single linear
least-squares fit.

Detection (Sect.~\ref{sec:detect}) and the moment estimators of
Eq.~(\ref{eq:moments13}) involve no differentiation of the reconstructed
profiles. Determining $\mbsq$ by profile fitting does, and differentiating $Z_1$
twice is sensitive to noise (Sect.~\ref{sec:blenddet}).
The numerical treatment of that noise requires care, since $Z_1$ and $Z_3$ do
not respond to it alike, so that a poor choice shifts the recovered $\mbsq$
instead of only affecting its precision.

The differentiation of the data
can, however, also be avoided altogether by imposing Eq.~(\ref{eq:tie}) within the
LSD calculation itself, which is the one step considered here that does call for a
change to the deconvolution code. For a trial value of $\gamma$ the line pattern matrix becomes
$\mathbf{M}_1 + \gamma\,\mathbf{M}_3\mathbf{D}_2$, where $\mathbf{D}_2$ is the
discrete second-derivative operator on the velocity grid. The problem then
remains linear in $Z_1$ at fixed $\gamma$, and $\gamma$ follows from a
one-dimensional scan in $\chi^2$. This adds no assumption beyond the one the
profile fit already makes, since both rest on Eq.~(\ref{eq:tie}) and thus on a
modulus nearly uniform over the regions where $B_\mathrm{z}$ does not vanish 
or a rotational broadening small compared with the width
of the local profile. It does impose the relation during the
deconvolution rather than after it, so that where the relation fails the
distortion is passed on to $Z_1$, and for a rapid rotator with a structured
surface the unconstrained solve is the safer choice.

One point deserves attention in the construction of the mask. The assumption
that the local residual profile scales with $d_i$ is not exact: saturated lines
have broader and flatter-cored profiles, so that $\partial I_i/\partial\lambda$
is not simply proportional to $d_i$. The resulting line-to-line variation of
profile shape is entirely non-magnetic. However, its signature in Stokes $V$
resembles that of the second-order term closely enough to contaminate $Z_3$, as
the low-field behaviour of the strong-line sample in Sect.~\ref{sec:tie} shows.
Restricting the mask to
lines of moderate central depth suppresses this undesired behaviour.

The truncation of the expansion at $n=3$ introduces a further systematic effect,
an underestimate of $\mbsq$ that grows with field strength. It can be calibrated
out by applying the identical procedure to synthetic spectra computed for the
parameters of the star, in the same spirit in which the interpretation of
first-order LSD profiles can be calibrated \citep{kochukhov:2014,kochukhov:2026}.

\section{Numerical validation}
\label{sec:val}

\subsection{Unblended local profiles}
\label{sec:unblended}

\subsubsection{Computational setup}
\label{sec:numexp}

In the first series of numerical experiments we tested the description developed above against local solar polarised spectra. 
A line list was extracted from VALD for solar parameters over the
interval 4000--7000\,\AA. The Stokes $I$ and $V$ profiles of each line were
computed individually with the {\sc synmast} code
\citep{kochukhov:2010a}, using a {\sc marcs} solar model atmosphere \citep{gustafsson:2008}, a
microturbulent velocity of 0.8\,\kms, and the full explicit Zeeman pattern of each
transition.
Of the 15\,850 lines in the extraction, the 5753 with a VALD central depth
above 0.1 and with a Land\'e factor available for both levels were
computed. A further 73 lines, whose absorption extended beyond the $\pm500$\,\kms\
synthesis window, were rejected, leaving 5680 lines.
The calculations were carried out for a depth-independent
magnetic field of $B$ = 0, 50, 100, 200, 500, 1000, 2000 and 5000~G at angles
$\theta = 0$, 30 and 60\degr\ between the field vector and the line of sight, all
at the centre of the stellar disk ($\mu = 1$).
Each profile was then
convolved with a Gaussian instrumental profile corresponding to
$R = 1.2\times10^5$, roughly the highest resolving power available to stellar
spectropolarimetry with instruments such as HARPSpol \citep{piskunov:2011} and PEPSI \citep{strassmeier:2003},
and resampled onto a common velocity grid with a step of 0.3\,\kms. 

The central depth $d$ of every line was measured from its zero-field Stokes $I$
profile after this convolution.
The list was divided at $d = 0.4$ into a sample of weak and moderately strong lines and a
sample of strong lines, referred to below as the weak-line and the
strong-line samples. Within each sample we discarded the lines whose
zero-field Stokes $I$ profile, rescaled to unit depth, departs from the median
profile of the sample by more than 2\% of the line depth in rms, evaluated over
the same velocity window as the fits below. This
step removes lines that violate the self-similarity assumption of LSD for reasons
that have nothing to do with the magnetic field, chiefly a formation depth
or broadening parameters
unlike those of the majority of the list. This selection retains 2697 of the 3299 lines with
$d<0.4$ and 854 of the 2381 lines with $d>0.4$, the smaller fraction in the
second case reflecting the flattened cores of saturated lines.

The two weight sets were formed from Eq.~(\ref{eq:weights}), 
with $\bar{g}$ and $G^{(3)}_1$ evaluated from the VALD Land\'e factors and quantum numbers through
Eqs.~(\ref{eq:gbar}) and (\ref{eq:g3}).
Each set of weights was normalised to a
reference line with $d = 0.2$, $\lambda_0 = 5000$\,\AA, $\bar{g} = 1.2$, and $G^{(3)}_1=\bar{g}^3=1.728$.
For each field strength the mean
profiles were then obtained by linear least squares from the model
(\ref{eq:model}) over the window $\pm25$\,\kms, which contains 166 pixels
and is several times wider than the profiles themselves. We retained either the
first component alone, equivalent to standard LSD, or both components.
The synthetic lines are unblended and share a common velocity grid, so the
problem decouples into an independent fit at each velocity point and isolates the
adequacy of the two-component model and of its weights from the conditioning of
the deconvolution itself.

\subsubsection{Accuracy of the two-component description}
\label{sec:accuracy}

\begin{figure*}
\fifps{\hsize}{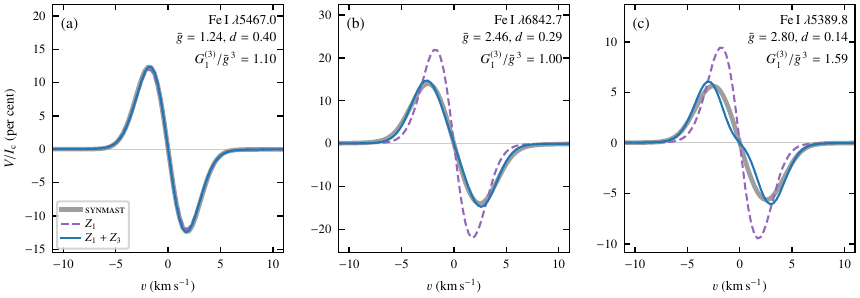}
\caption{Synthetic Stokes $V$ profiles of three individual lines of the weak-line
sample at $B = 1$~kG (thick grey lines), compared with the single-component
(dashed lines) and two-component (solid lines) LSD models. The mean profiles entering the
models are those recovered from the sample as a whole, and the contribution of
each line is fixed by its own $d$, $\lambda_0$, $\bar{g}$ and $G^{(3)}_1$.}
\label{fig:lines}
\end{figure*}

\begin{figure}
\fifps{\hsize}{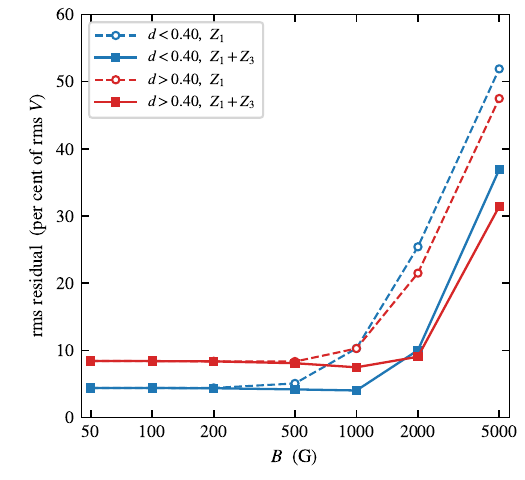}
\caption{Rms difference between the synthetic Stokes $V$ profiles and the LSD
model, pooled over all lines of a sample and all velocity points of the
fitting window, as a percentage of the rms of Stokes $V$ over the same points. Open symbols and dashed lines
correspond to standard, single-component LSD; filled symbols and solid lines to
the two-component description of Eq.~(\ref{eq:model}). Blue and red refer
respectively to the samples of weak and strong lines defined in
Sect.~\ref{sec:numexp}.}
\label{fig:residual}
\end{figure}

Figure~\ref{fig:lines} shows three lines of the weak-line sample at 1~kG together with the single-component and two-component LSD models. Nothing
in either model has been fitted to these specific lines: the mean profiles come from the
sample as a whole, and each line contributes with the weights of
Eq.~(\ref{eq:weights}) evaluated from its own parameters.

For $\bar{g} = 1.24$ (Fig.~\ref{fig:lines}a), close to the median of the mask, the two descriptions are
indistinguishable from the synthetic profile and from each other. However, for
$\bar{g} = 2.46$ (Fig.~\ref{fig:lines}b) the single-component model overestimates the amplitude by about
one half and distorts the shape. Adding the second component brings it
back, from 9.4 to 1.1\% of the line amplitude in rms. The third line
(Fig.~\ref{fig:lines}c) has a comparable $\bar{g}$ but a strongly
anomalous splitting pattern, $G^{(3)}_1/\bar{g}^{\,3} = 1.59$. Here the
second component repairs most but not all of the discrepancy, 10.9 to 3.6\%.

That the first-order error is an overestimate for these two lines is not accidental. The
weight $w^{(1)}$ is linear in $\bar{g}$, whereas the
amplitude of Stokes $V$ grows more slowly than linearly once the splitting ceases
to be small compared with the line width. A single mean profile can only
average that departure over the mask, so it overshoots for a line of large
$\bar{g}$ and undershoots for the small-$\bar{g}$ majority. The second-order term is the leading correction for
that saturation.

\begin{table}[!t]
\caption{Median rms difference between the synthetic Stokes $V$ profile of an
individual line and the single- or two-component LSD model.}
\label{tab:gbar}
\centering
\begin{tabular}{lrrr}
\hline\hline
\noalign{\smallskip}
$\bar{g}$ & $N$ & $Z_1$ & $Z_1+Z_3$ \\
\noalign{\smallskip}
\hline
\noalign{\smallskip}
$<0.5$ &  126 & 1.84 & 0.71 \\
0.5--1.0 &  625 & 1.59 & 0.56 \\
1.0--1.5 & 1390 & 1.06 & 0.46 \\
1.5--2.0 &  439 & 0.68 & 0.36 \\
2.0--2.5 &   91 & 1.96 & 0.38 \\
2.5--3.2 &   23 & 4.94 & 0.45 \\
\noalign{\smallskip}
\hline
\end{tabular}
\tablefoot{The rms for $Z_1$ and $Z_1+Z_3$ models is given as a 
percentage of the peak-to-peak amplitude of
that line, in bins of effective Land\'e factor. $N$ is the number of lines
in the bin. Weak-line sample, $B = 1$~kG.}
\end{table}

\begin{figure*}[!t]
\fifps{\hsize}{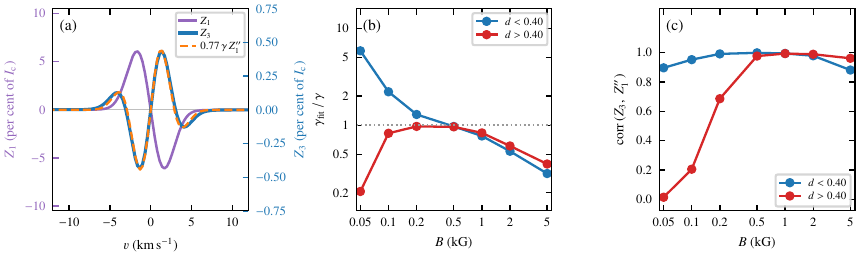}
\caption{\textbf{(a)}: the two mean profiles recovered from the weak-line
sample at $B = 1$~kG. $Z_1$ refers to the left axis and $Z_3$ to the right one.
The dashed line is $Z_1''$ scaled by the coefficient that best reproduces $Z_3$,
its value relative to $\gamma$ predicted by Eq.~(\ref{eq:gamma})
being given in the legend. \textbf{(b)}: that coefficient in units of
$\gamma$, which for a homogeneous field is the ratio of the recovered to the true
mean square field modulus, against field strength for the two line samples. The
dotted line marks the exact value. \textbf{(c)}: the linear (Pearson)
correlation coefficient between $Z_3$ and $Z_1''$ over the velocity points of the
fitting window.}
\label{fig:profiles}
\end{figure*}

These three lines are representative. Over the whole weak-line sample at 1~kG the
median rms difference for an individual line falls from 1.17 to 0.47\% of
its amplitude, and the ninetieth percentile from 2.04 to 0.94\%.
Table~\ref{tab:gbar} shows the median rms in bins of $\bar{g}$. The
single-component value is not a monotonic function of $\bar{g}$, the rise above
$\bar{g} \simeq 2$ being linked to the failure of the weak-field approximation.
The two-component values lie between 0.36
and 0.71\% across the entire range, so the second component decreases the dependence 
of the accuracy on $\bar{g}$.

The distortion that the single-component fit imposes on the mean profile can
also be measured. For a homogeneous field the first-order profile is
known: it is $kcB_\mathrm{z}\,\partial\mathcal{I}/\partial\varv$.
Recovered on its own, $Z_1$ reaches 0.97, 0.86 and 0.60 of
that amplitude at 500~G, 1~kG and 2~kG, and departs from its shape by 0.4, 1.9
and 5.7\% of it in rms. Recovered jointly with $Z_3$, it
gives 1.01, 0.99 and 0.83, and 0.1, 0.3 and 2.8\%.
The first moment is unaffected either way (Sect.~\ref{sec:moments2}).

Figure~\ref{fig:residual} condenses this comparison into one number for
each line group and field strength. The difference between the synthetic profile and
the model is formed at every velocity point of every line of the sample, and its
rms is divided by the rms of the synthetic Stokes $V$ over the same points. The
ordinate is thus the fraction of the circular polarisation signal that the
model fails to reproduce. Every line enters with its own amplitude instead of
being normalised to it, so the measure follows the strongest signatures of the
mask. It is not comparable with the per-line percentages of Table~\ref{tab:gbar},
which are referred to the peak-to-peak amplitude of each line and are
some seven times smaller for that reason alone.

Below about 200~G the two descriptions are indistinguishable and the residual is
independent of field strength. It measures the line-to-line variation of profile
shape that survives in the mask. This floor sets the accuracy with which any LSD
model built on a single common profile can reproduce the individual profiles,
4.4\% for the weak-line
sample and 8.4\% for the strong-line sample. From 500~G onwards the
single-component residual leaves this floor and grows rapidly, reaching 10\%
at 1~kG and 25\% at 2~kG for the weak-line sample. This growth is
the failure of the weak-field approximation, and it appears at field strengths
that are typical of active stars.

With the second component the failure sets in at much stronger fields.
At 1~kG the two-component residual is still at the
field-independent floor,
and at 2~kG it reduces the single-component residual by three fifths. At
5~kG both descriptions fail, as expected from the truncation of a series in
$\zeta$ that is no longer small.

These two line samples differ in the way anticipated in Sect.~\ref{sec:practical}. For
the strong lines the floor is twice as high, because a common profile scaled by
the central depth is a poorer description of saturated lines. The
two-component residual remains at 7.5\% at 1~kG against 4.0\% for the weak
lines, so the description of the strong-line sample is limited by the non-magnetic floor rather than
by the field. This argues for restricting the mask to lines
of moderate central depth if the second component is to be exploited
quantitatively.

\subsubsection{Relation between the two mean profiles}
\label{sec:tie}

The residuals of Fig.~\ref{fig:residual} establish that a second LSD component with
the weights $w^{(3)}$ improves the fit to synthetic data, but not that it carries the information claimed
in Sect.~\ref{sec:moments2}. That claim requires the relation (\ref{eq:tie}) between
the two profiles, and the spectra used here test it under the conditions for
which it is exact, namely a single point of the stellar surface with one value of the
field modulus.

Figure~\ref{fig:profiles}a shows the two profiles recovered
from the weak-line sample at 1~kG, together with $Z_1'' \equiv
\partial^2Z_1/\partial\varv^2$ scaled by the coefficient that best reproduces
$Z_3$. The second component is some 7\% of the first in amplitude, and
it is clearly not an independent function of velocity. The linear (Pearson) correlation
coefficient between $Z_3$ and $Z_1''$, evaluated over the velocity points of the
fitting window, is 0.995. This is the property that makes the field strength
accessible. Had $Z_3$ been free to take any shape, its amplitude would have been
degenerate with that shape, and no combination of the two profiles would have
determined $B$.

The scale factor between $Z_3$ and $Z_1''$ is a measurement of $\gamma$, and thus of $\mbsq$ through
Eq.~(\ref{eq:gamma}). Since the field is homogeneous here, the ratio
$\gamma_{\rm fit}/\gamma$ plotted in Fig.~\ref{fig:profiles}b is the ratio of the
recovered to the true $\mbsq$, and its square root is the corresponding factor on the
field modulus. At 500~G the ratio is within 4\% of unity for both
samples, so the estimator is unbiased where the expansion is accurate and the
second component is well determined. However, towards stronger fields it falls
monotonically, to 0.77 at 1~kG and 0.54 at 2~kG for the weak-line sample,
corresponding to underestimates of the field modulus by 12 and 27\%. This
is the effect of truncating the series (\ref{eq:hv}) at $n=3$, the next term
carrying the opposite sign and removing part of the third-order amplitude. Being a
smooth and monotonic function of field strength, it can be calibrated in the
manner described in Sect.~\ref{sec:practical}.

Towards weaker fields the ratio also departs from unity, upwards for the
weak-line sample and downwards for the strong-line one. Here the cause is not the
expansion but the amplitude of the second component itself, which falls as $B^2$
until it is no longer separable from the line-to-line profile variation that sets
the floor of Fig.~\ref{fig:residual}. Figure~\ref{fig:profiles}c makes the transition
explicit: the correlation coefficient between $Z_3$
and $Z_1''$ exceeds 0.99 from 200~G to 2~kG for the weak-line sample, but falls to
0.90 at 50~G. For the strong-line sample, whose floor is twice as high, it
reaches only 0.69 at 200~G and 0.01 at 50~G. What remains of $Z_3$ in that
regime is residual profile variation rather than signal, and the coefficient
fitted to it is not a measurement of the field. 

At the same time, between 500~G and 2~kG the two samples return values of
$\gamma_{\rm fit}$ that agree to better than 12\%, and to 6\% on
the field modulus, although their residuals in Fig.~\ref{fig:residual}
differ by a factor of two. The quality with which the model reproduces the
individual profiles and the accuracy of the field strength derived from it are
thus distinct questions, and the second is the more forgiving of the two.

The tests above refer to a field directed along the line of sight, but they
are not restricted to it. Repeating the analysis at $\theta = 30^\circ$ and
$60^\circ$, computed at the same $\mu = 1$, changes the residuals of
Fig.~\ref{fig:residual} by no more than 0.3 percentage points and gives
$\gamma_{\rm fit}/\gamma = 0.774$, 0.777 and 0.783 at 1~kG for $\theta = 0^\circ$,
$30^\circ$ and $60^\circ$. The corresponding spread in the field modulus is below
0.7\%, against a truncation bias of 12\% at the same field strength, so the
weak-line results may be read as applying to a field of any orientation.

\subsection{Blended disk-integrated spectra}
\label{sec:blend}

\subsubsection{Computational setup}
\label{sec:blendsetup}

\begin{figure*}[!t]
\fifps{\hsize}{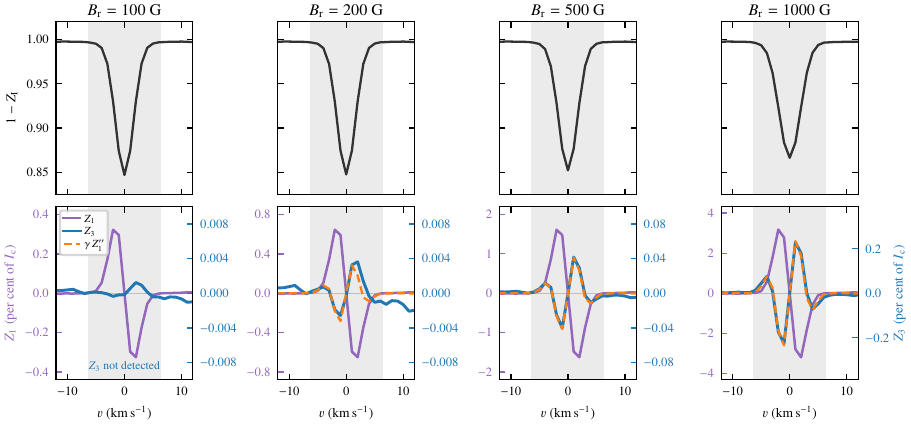}
\caption{Mean profiles recovered from the blended disk-integrated spectra, for a
radial field of the strength given above each column. \textit{Upper panels}: the mean
intensity profile.
\textit{Lower panels}: $Z_1$ on the left axis and $Z_3$ on the right one.
The dashed curve is $Z_1''$ scaled by the fitted $\gamma$. 
Shading marks the
velocity interval over which the moments and the detection test are evaluated. At
100~G the second component fails the false-alarm probability test and no scaled derivative is
drawn.}
\label{fig:blendprof}
\end{figure*}

The synthesis of one line at a time and the restriction to disk centre are both
removed in the second series of numerical experiments. All 15\,850 lines of the same VALD extraction as in Sect.~\ref{sec:numexp}
were computed together, so that every line appears with appropriate blends, and the
specific intensities were integrated over the visible hemisphere on seven
limb-angle nodes to give Stokes $I$ and $V$ flux spectra of the whole
4000--7000\,\AA\ interval. The field is radial and of constant modulus,
$B$ = 100, 200, 500, 1000 and 2000~G, and the star does not rotate, so that
Eq.~(\ref{eq:tie}) is exact. The spectra were convolved to $R = 1.2\times10^5$
and sampled at 0.5\,\kms.
Mean profiles were obtained with the \textsc{iLSD} code
\citep{kochukhov:2010a} without any modification, on a velocity grid
from $-25$ to $+25$\,\kms\ with a 1\,\kms\ step.
For a radial
field of constant modulus the disk averages of Eq.~(\ref{eq:diskav}) are known
analytically, $\mbz = B\langle\mu\rangle$ and $\mbsq = B^2$, where
$\langle\mu\rangle = 0.715$ is the cosine of the limb angle weighted by the continuum
intensity of the model atmosphere and averaged over the mask.

Blending calls for two selection steps that unblended profiles were not subjected to, both made from
the intensity spectrum and the line list alone. The first rejects the pixels in
which a broad feature depresses the continuum. The local pseudo-continuum, taken
as the running maximum of Stokes $I$ over 30\,\kms, is several times wider than a
photospheric line and therefore remains near unity between the narrow lines even
in the crowded blue, where its median is 0.97, and the pixels at which it is
depressed by more than 0.10 were rejected. This amounts to 6.7\% of the interval
in 45 stretches, and removes all four Balmer lines, both Na~D components,
the three Mg~b lines and other strong metal lines.
The second rejects the pixels to which lines not retained in the final mask contribute.
Every line of the extraction with $d > 0.05$ that does not enter the mask has the
interval within 2.5 FWHM of its centre removed, 42.5\% of the range, and the
pseudo-continuum test is repeated at a depression of 0.03, which removes 17.2\%
of the range. The pseudo-continuum was evaluated once, on
the zero-field intensity spectrum, and held fixed across the grid, the Zeeman
broadening of Stokes $I$ for $B=2$~kG being negligible on the 30\,\kms\ scale of the running
maximum.

The Stokes $I$ shape tolerance of Sect.~\ref{sec:numexp} cannot be applied to a
blended spectrum, which does not provide the profile of a line on its own. Its
place is taken by a criterion on the line width, a mask line being rejected when
the FWHM of its profile normalised to unit depth departs by more than 15\% from
the running median at its central depth. These widths come from the synthesis of
each line in isolation, which for a real star means a spectrum computed for its
parameters, as the calibration of Sect.~\ref{sec:practical} requires in any case.
With the same depth limit $d < 0.4$ the width criterion rejects 1355 of 5913
lines and the pixel rejection a further 2964, leaving a mask of 1594 lines and
53.5\% of the pixels. With the depth cut alone the residual at 1~kG is 93\% of
the rms of Stokes $V$ and the correlation of $Z_3$ with $Z_1''$ only 0.90, against
20\% and 0.994 for the adopted selection.

Photon noise was added to the input Stokes $V$ spectrum as Gaussian deviates of
standard deviation $1/(S/N)$ in units of the continuum, per 0.5\,\kms\ pixel,
for $S/N=200$, 500, and 1000. Every signal-to-noise ratio quoted below refers
to the input Stokes $V$ spectrum and not to the mean profile, which is more precise by the
multiplex factor of the mask.

\subsubsection{Mean profiles and the detection limit}
\label{sec:blenddet}

Figure~\ref{fig:blendprof} shows the profiles recovered at four field strengths
from the noise-free spectra.
The second component rises from 0.6\% of the first in amplitude at 200~G to 7\%
at 1~kG, the latter being the ratio found at the same field strength for
unblended lines (Sect.~\ref{sec:tie}). Where it is detected, $Z_3$ again
reproduces $Z_1''$, the correlation coefficient over the velocity grid being
0.980, 0.994 and 0.982 at 500~G, 1 and 2~kG.

Whether the second component is present at all was decided by the $\chi^2$
false-alarm probability test widely used for LSD profiles \citep{donati:1992}, applied over the
velocity interval in which the moments are taken. $Z_3$ is a definite detection
from 200~G upwards, with a false-alarm probability of $<$\,$10^{-5}$ at 200~G,
and it is not detected at 100~G, where the correlation with $Z_1''$ is 0.28.
Below the detection threshold $\mbsq$ is undefined: the coefficient fitted at
100~G exceeds the prediction of Eq.~(\ref{eq:gamma}) by 59\%, and it may equally
come out negative, which has no square root at all.

Blending raises the floor of Fig.~\ref{fig:residual} without changing what lies
above it. The rms residual of the two-component fit at 1~kG is 20\% of the rms of
Stokes $V$ over the retained pixels, against 4.4\% for the unblended weak-line
sample, the difference being the circular polarisation of the lines left out of
the mask, which no mean profile can reproduce. The low-field limit moves up by
about a factor of two, $Z_3$ still being correlated with $Z_1''$ at 0.95 at 100~G
in the unblended experiment. The longitudinal field is unaffected by any of
this, and is recovered at 0.969, 0.964, 0.953 and 0.925 of $B\langle\mu\rangle$
at 100, 500, 1000 and 2000~G.

Evaluation of the moments does require a restricted velocity window here, which the noise-free unblended experiment did
not. An incomplete mask produces a pedestal in the mean intensity profile
and leaves the wings of
$Z_1$ and $Z_3$ with structure that $\varv$ and $\varv^3$ weight heavily. The
continuum of the Stokes $I$ LSD profile
was therefore subtracted before obtaining $W_I$, and all
moments were calculated over a single interval, set at 1.6 times the half-width of
$Z_1$ at 5\% of its peak. This window, shown in Fig.~\ref{fig:blendprof}, changes from $\pm6.4$\,\kms\ for weak-field profiles to $\pm8.0$\,\kms\ at 2~kG where Zeeman broadening becomes noticeable. 

The coefficient $\gamma$ behaves as it did on unblended lines. Fitting $Z_3$
against $Z_1''$ returns 1.071, 0.975, 0.869 and 0.731 of the true field modulus
at 200~G, 500~G, 1 and 2~kG,
reproducing the disk-centre values of Sect.~\ref{sec:tie} to better than
1.5\% in field modulus at the three strongest field values.
The truncation bias is thus a property of the Zeeman expansion and of the
mask rather than of the spectra, and blending raises the noise floor of the
measurement while leaving its systematic behaviour untouched.

With noise added to the spectra, the field strength at which the second component becomes available
depends on $S/N$ as well. Over 200 realisations $Z_3$ is
detected in every one at $S/N = 200$ for $B \ge 500$~G, so that in the regime of
interest for active cool stars the measurement is limited by systematic effects and not by photon
noise. At 200~G it is found in 3\% of the realisations at $S/N = 500$ and in 40\%
at $S/N = 1000$, and at 100~G in none of them. Where detection is found in a small number of realisations, 
the resulting $Z_3$ tends to be spuriously high.

Differentiation of $Z_1$ and the regularisation of the deconvolution both act on
the second component more strongly than on the first. 
We found that a five-point second-derivative operator gives the best results,
lower-order schemes underestimating the curvature of $Z_1$.
Regularisation is the more serious of the two, since $Z_3$ is
essentially $Z_1''$ and carries more of its power at high frequency, so that a
penalty on the first differences of both profiles suppresses it
disproportionately. The second component should therefore be
recovered without regularisation, and where noise has to be traded for resolution
the velocity grid should be coarsened instead, which weights the two profiles alike.

\subsubsection{Field strength estimation}
\label{sec:blendfill}

\begin{figure}
\fifps{\hsize}{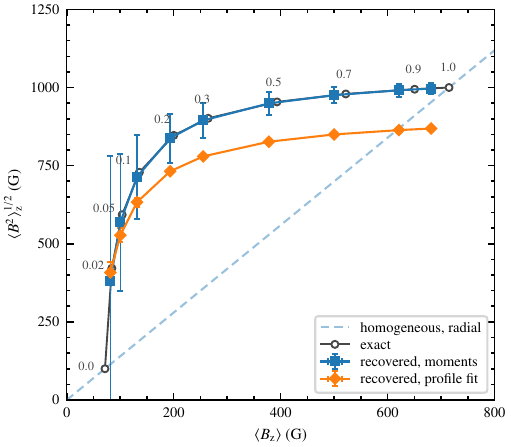}
\caption{Mean square field modulus against mean longitudinal field for the
blended spectra at $S/N = 500$, for a surface a fraction $f$ of which carries a radial field of
1~kG while the remainder carries 100~G. Numbers along the exact track give $f$.
Open circles show the exact values of Eqs.~(\ref{eq:diskav}) and
(\ref{eq:bsqz}), squares the moment ratio of Eq.~(\ref{eq:moments13}) and
diamonds the value obtained by fitting $Z_3$ against $Z_1''$.
The dashed line is the locus of a homogeneous radial field,
$\mbsq^{1/2} = \mbz/\langle\mu\rangle$.
At $f = 0$ the second component is not detected and no recovered value is plotted.}
\label{fig:blendfill}
\end{figure}

The separation of field strength from filling factor was tested on these
spectra. For circular polarisation the emergent profile of a surface covered
to a fraction $f$ by a field $B_2$, the remainder carrying $B_1$, is the
area-weighted sum
\begin{equation}
V(\varv) = f\,V_{B_2}(\varv) + \left(1-f\right)V_{B_1}(\varv)\,,
\label{eq:twocomp}
\end{equation}
which holds for the disk-integrated flux spectrum exactly as it does for a single
surface element. The spectrum of a surface with $B_2 = 1$~kG and $B_1 = 100$~G is thus obtained as a weighted mean of two disk-integrated Stokes $V$ spectra. The exact
averages are $\mbz = \left[fB_2+(1-f)B_1\right]\langle\mu\rangle$ and
$\mbsq = \left[fB_2^3+(1-f)B_1^3\right]/\left[fB_2+(1-f)B_1\right]$, the second
independent of the disk weighting because the modulus is uniform over each
component.

Figure~\ref{fig:blendfill} follows the track from $f = 0$ to unity, with both
estimators of $\mbsq$ formed from a single noise realisation at $S/N = 500$ and
their errors propagated from the covariance matrix of the LSD solution. The
profile fit was in addition corrected for the noise in its own regressor.
The moment ratio of Eq.~(\ref{eq:moments13}) is accurate to between
0.3 and 4\% at every $f \ge 0.02$, and its departure from the exact value is
within its own error bar at every point. The profile fit lies 11 to 14\% low for
$f \ge 0.05$, almost independently of $f$, so
that both tracks run parallel to the exact one and well above the locus of a
homogeneous radial field. The independence of $\mbsq$ from the filling factor
is thus established under blending, disk integration and finite noise,
allowing one to separate $B$ from $f$. At $f = 0$ the second
component is not detected, which is the low-field floor of
Sect.~\ref{sec:blenddet} and not a failure of the estimator.

That the moment ratio is unbiased where the profile fit is not follows from
Eq.~(\ref{eq:momzero}). The third moment of the $n$-th derivative of the local
profile vanishes for every odd $n$ above 3, so that
$\int\varv^3Z_3\,{\rm d}\varv$ is untouched by the terms of the expansion
(\ref{eq:hv}) that the truncation neglects, whereas a fit of $Z_3$ to $Z_1''$
absorbs them into $\gamma$. The truncation bias of Sect.~\ref{sec:tie} is a
property of the estimator and not of the second component. The moment
ratio therefore requires no calibration, which is the one respect in which it is
preferable to the better conditioned fit.

What the moment ratio costs is the wings of $Z_3$, which the $\varv^3$ weighting
samples heavily, 93\% of the integral over the whole velocity grid coming from
$|\varv| > 8$\,\kms. On blended spectra those wings carry a residual that scales
as $B$ rather than as $B^3$ and is anticorrelated with $Z_1$,
the unmodelled circular polarisation of the lines outside the mask distributed
between the two components by the fit. Since the signal grows as $B^3$ and the
contamination as $B$, the resulting bias falls as $B^{-2}$, from 47\% at 200~G to
5\% at 2~kG. A cleaner mask removes it where a correction does not. Widening the
interval rejected around each contaminant line from 1.5 to 2.5 FWHM costs 367 mask lines
and 9\% of the pixels, changes the profile fit by less than 0.005, and takes the
moment from 1.24, 1.07 and 1.08 to 1.09, 1.00 and 1.05 of the true modulus at
500~G, 1 and 2~kG.

Two limitations of the method deserve emphasis. The calibration of the
profile fit depends on the distribution of field strength. With
components of 100~G and 2~kG the recovered fraction runs from 0.56 to 0.74 along
the track, and at $f = 0.05$ it falls 17\% below what a homogeneous 2~kG
calibration predicts, so that a calibration should be computed for a plausible
distribution of field strengths rather than for a single representative value. The depth
limit cannot be relaxed to gain lines either. For a homogeneous field the
recovered modulus is insensitive to it up to $d = 0.6$, whereas at $f = 0.05$,
where the second component is a small fraction of the first and the floor is what
limits the measurement, it falls from 0.89 of the exact value at $d < 0.4$ to
0.75 at $d < 0.6$ and is not monotonic in between, 0.69 at $d < 0.5$. 

\section{Summary and conclusions}
\label{sec:concl}

In this paper we have developed a second-order extension of least-squares
deconvolution which gives access to the strength of a stellar magnetic field, and
not merely to the longitudinal flux it carries, from circular polarisation alone.
It rests on an exact property of the Zeeman effect. The circular polarisation
absorption coefficient of a spectral line is
a series in odd wavelength derivatives of the intensity profile, with
coefficients given by the moments of its Zeeman pattern. Retaining the first
two terms of that series and integrating over the visible stellar hemisphere
yields a description of the multiline Stokes $V$ spectrum in which every line
enters through two weights. The first is the $d\lambda_0\bar{g}$ of standard LSD,
the second a weight $d\lambda_0^3G^{(3)}_1$ formed from the third moment of the
splitting pattern of that line. The two mean profiles recovered with these
weights are built from the same local profile, but they are averaged over the
stellar surface with $B_\mathrm{z}$ in one case and with $B^2B_\mathrm{z}$ in the
other. It is that difference of weighting which carries information the
longitudinal field does not. The ratio of their moments defines a new magnetic observable, 
the mean square field modulus $\mbsq$, which allows one to separate the
strength of the field from the surface area that it occupies. This is the multiline analogue of the Stokes $V$ line-ratio
technique introduced for the Sun by \citet{stenflo:1973}, and it calls for no new
observations and, in its basic form, for no modification of existing multiprofile
LSD codes.

This description has been tested against polarised spectrum synthesis, first for
individual lines computed at the centre of the stellar disk and then for blended
flux spectra covering the entire optical region. The second Stokes $V$ component $Z_3$ removes
most of the error that a single mean profile makes for lines whose Stokes $V$ profiles begin to depart from 
the weak-field approximation. It is moreover not an independent
function of velocity, following the second derivative of the first-order profile $Z_1$ with a
correlation exceeding 0.99 between 200~G and 2~kG for lines of moderate central
depth. It is detected from about 200~G in noise-free spectra and, over the whole
range of field strengths of interest for active cool stars, at the
signal-to-noise ratios typical of modern spectropolarimetry. Two estimators of $\mbsq$
follow from it, and they are limited in opposite ways. The moment ratio
of the two profiles is unaffected by the truncation of the series and
therefore requires no calibration, but its $\varv^3$ weighting amplifies
noise and unmodelled polarisation contamination in the far profile wings. Fitting
$Z_3$ against $\partial^2Z_1/\partial\varv^2$, or imposing that relation within
the deconvolution itself, is far better conditioned, at the cost of an
underestimate that grows with field strength and has to be calibrated on
synthetic spectra. Both estimators return a field modulus independent of the filling factor
under blending, disk integration and photon noise.

Appendix~\ref{sec:app} applies the method to archival ESPaDOnS observations of
two cool Ap stars (\equ\ and \aql) and two active M dwarfs (\leo\ and \lac). Four spectra of each target were
analysed, selected without regard to rotational phase and without any attempt to
exploit the full data sets available for these well-studied objects, so that the
results are representative rather than definitive. $Z_3$ is nevertheless detected
in all sixteen spectra, and the constrained deconvolution is the only estimator
that yields a precise value for every star, the moment ratio remaining usable on
the two M dwarfs and on \equ. For the Ap stars the recovered modulus is
compatible with the field strength known from resolved or modelled Zeeman
splitting. For the M dwarfs it falls close to the moduli measured by Zeeman
broadening and far above the strength of the large-scale field that ZDI
reconstructs from the same data. A definitive analysis of any of these
stars requires the identical procedure to be applied to synthetic spectra
computed for its own parameters.

The value of this diagnostic is not the same everywhere in the H-R diagram. For
early-type stars with fossil fields it adds comparatively little. Their
magnetic geometries are globally organised and largely free of the small-scale
component to which the second-order profile responds, and their field moduli are
already accessible from the Zeeman splitting or intensification of the intensity
spectrum. Cool active stars are the opposite case. Their fields are multi-scale,
most of the magnetic energy resides at spatial scales whose polarisation
signatures cancel, and the circular polarisation that survives that cancellation
refers to a larger-scale component whose local strength has never been
observable. The relation between polarimetric observables and the moduli obtained
from Zeeman broadening or intensification has consequently remained uncertain,
undermining ZDI inversion results \citep{kochukhov:2026}.
The second mean Stokes $V$ profile supplies the missing local field strength
directly, giving for the first time the strength of the field in the regions from
which the circular polarisation emerges. Its diagnostic weight relative to the
first-order one grows as $\lambda^2$, so that the near-infrared spectropolarimeters
now in operation are an order of magnitude more sensitive to it than the optical data
used here.

\begin{acknowledgements}
Support from the Swedish Research Council (grant agreement no. 2023-03667) and the Swedish National Space Agency is acknowledged.
\end{acknowledgements}

\begin{appendix}
\nolinenumbers

\section{Representative applications}
\label{sec:app}

\subsection{Cool Ap stars \equ\ and \aql}
\label{sec:apstars}

The second-order LSD analysis was applied to two cool (\teff\ = 7500--7800~K) Ap stars, both of them
slow rotators with sharp lines, for which Eq.~(\ref{eq:tie}) is expected to hold,
and both with a field modulus known independently from the intensity spectra.
\equ\ (HD\,201601) has a rotation period of the order of a
century \citep{bychkov:2016} and a modulus measured from the resolved Zeeman
doublet of Fe\,{\sc ii} 6149\,\AA\ that decreased from 3.80~kG in 2011 to
3.58~kG in 2017 \citep{jarvinen:2024}. \aql\ (HD\,176232) has a rotation period exceeding 12 yr \citep{sikora:2019a} and
shows no resolved Zeeman splitting in the optical. Its field modulus of 1.4--1.5~kG was
derived from detailed line profile modelling and from near-infrared observations
\citep{kochukhov:2002,leone:2003a}. Four independent ESPaDOnS Stokes $V$ observations of
each star were retrieved from the PolarBase archive \citep{petit:2014}, forming
two pairs taken 40~min apart and separated by 14~months for \equ\ and a sequence spanning
five years for \aql, at $S/N$ = 670--990 and 421--662 per pixel.

Line masks were extracted from VALD for model
atmospheres with the parameters and the chemical abundances of each star
\citep{shulyak:2013,nesvacil:2013}. After removing the regions affected by the Balmer lines, the telluric bands
and the pixels rejected by the pseudo-continuum test of
Sect.~\ref{sec:blendsetup}, there remained 6820 lines for \equ\ and 10\,382 for \aql\ with central depths above 5\% of the continuum.
The central depths, the least secure
ingredient of a mask built from a model, were then adjusted with the iterative
line-weight refinement of \textsc{iLSD} from the observed
intensity spectrum alone. Three passes lower the reduced chi-square of the Stokes
$I$ fit by a factor of 5 for \equ\ and 10 for \aql, and improve the Stokes $V$
fit by 19 to 33\%. Restricting the mask to
$d < 0.4$ on these adjusted depths, as in Sect.~\ref{sec:blendsetup}, leaves 3352 and 4388 lines.

The mean profiles were reconstructed on a 1.5\,\kms\ grid, close to the mean pixel
spacing of the observed spectra, and the moments were calculated over the $\pm15$\,\kms\ window that
contains both profiles. 
At this sampling the unconstrained fit of $Z_3$ against $Z_1''$ cannot be carried
out at all, its regressor being dominated by noise, so that the constrained
deconvolution of Sect.~\ref{sec:practical} was used for both stars. The moment
ratio of Eq.~(\ref{eq:moments13}) remains available for \equ, where it gives
3344~G on average, 14\% below the constrained fit, a difference of less than
twice its own uncertainty of about 315~G. These moment measurements scatter by 7\%
between spectra taken 40~min apart, against 0.3\% for the fit.
For \aql\ the moment ratio yields no usable result, its error exceeding the
value itself. This is due to the
$\varv^3$ weighting reaching the wing contamination of
Sect.~\ref{sec:blendfill} in a spectrum far more crowded than the synthetic
one.

\begin{figure}[!t]
\figps{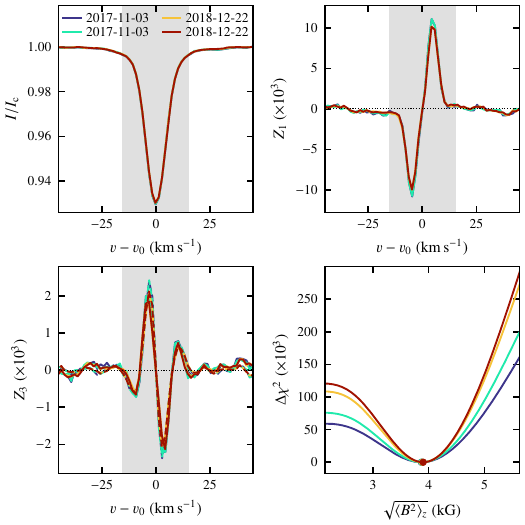}
\caption{Second-order LSD analysis of the four ESPaDOnS observations of
\equ. \textit{Upper panels}: the Stokes $I$ LSD profile and the
first-order profile $Z_1$. \textit{Lower left}: the second-order profile $Z_3$ from the
unconstrained two-component deconvolution (solid lines) and the constrained solution
$\gamma Z_1''$ (dashed lines). \textit{Lower right}: $\Delta\chi^2$ of the constrained fit to the
observed Stokes $V$ spectrum as a function of the corresponding field strength,
the circle marking its minimum. Shading marks the velocity interval over which
the moments are evaluated.}
\label{fig:gamequ}
\end{figure}

The second Stokes $V$ component is detected in every spectrum of \equ\
(Fig.~\ref{fig:gamequ}), reaching 21\% of the amplitude of
$Z_1$, the constrained and the unconstrained solutions coinciding. The four
spectra give $\mbsq^{1/2}$ = 3883~G (Table~\ref{tab:apstars}) with a scatter of 32~G, or 0.8\%,
compared with 24~G of formal uncertainty. A centred dipole yields
$\mbsq^{1/2}/\langle B\rangle$ between 1.03 and 1.10 whatever its orientation,
so that this measurement implies \bm\ = 3.5--3.8~kG at the epoch of the
observations, where the modulus of the star was about 3.6~kG.

\begin{figure}[!t]
\figps{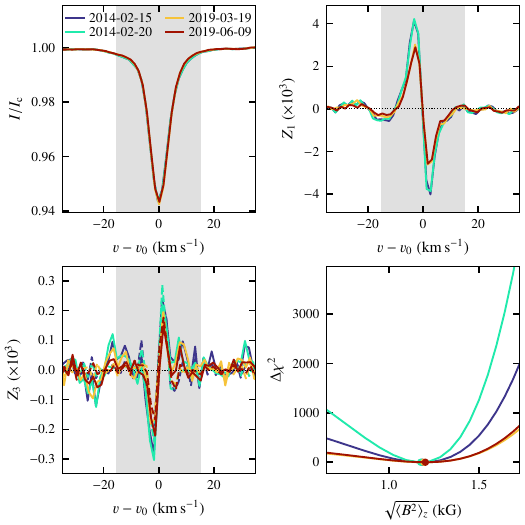}
\caption{Same as Fig.~\ref{fig:gamequ} for the four ESPaDOnS observations of \aql.}
\label{fig:aql}
\end{figure}

For \aql\ the second component is also detected in all four observations. It amounts to 6.8\% of the amplitude of $Z_1$, and the
unconstrained $Z_3$ is dominated by noise (Fig.~\ref{fig:aql}), so that only the
constrained solution is suitable for a quantitative measurement. The four
spectra give 1190~G with a scatter of 26~G and a typical uncertainty of 55~G,
implying \bm\ = 1.09--1.16~kG. Between 2014 and 2019 \bz\ fell by 77~G, about twenty
times the uncertainty of that difference, whereas $\mbsq^{1/2}$ moved by 17~G
between the two epochs, a third of its own error bar. The star rotated in that
interval, altering which part of the surface was seen and with it the
longitudinal field, while the modulus over the visible hemisphere remained roughly the
same, yielding a stable $\mbsq$ measurement.

The moduli obtained from the constrained fit are uncalibrated. 
The truncation bias of
Sect.~\ref{sec:blenddet}, measured on the solar spectrum at the field strength of \aql, brings its
implied modulus to 1.24--1.32~kG, whereas the field strength of \equ\ places it outside this calibration.
The bias is
a saturation effect and thus a property of the line strengths of the star, so
that its calibration has to be established on synthetic spectra computed for the
star itself, as Sect.~\ref{sec:practical} suggested. That both uncalibrated fit values
lie within 20\% of independently measured moduli indicates that this correction
is small for cool Ap stars.

\begin{table}
\caption{Mean longitudinal magnetic field and mean square field modulus from second-order LSD analysis.}
\label{tab:apstars}
\centering
\small
\setlength{\tabcolsep}{3pt}
\begin{tabular}{lr r@{$\,\pm\,$}l r@{$\,\pm\,$}l r@{$\,\pm\,$}l}
\hline\hline
 & \multicolumn{1}{c}{HJD} & \multicolumn{2}{c}{\bz\ (G)} & \multicolumn{4}{c}{$\mbsq^{1/2}$ (G)} \\
UT date & \multicolumn{1}{c}{$-2450000$} & \multicolumn{2}{c}{moment} & \multicolumn{2}{c}{moment} & \multicolumn{2}{c}{fit} \\
\hline
\multicolumn{8}{l}{\equ} \\
2017-11-03 & 8060.67789 & $-846$ & 8 & 3715 & 310 & 3869 & 24 \\
2017-11-03 & 8060.70457 & $-856$ & 8 & 3454 & 311 & 3875 & 24 \\
2018-12-22 & 8474.68271 & $-813$ & 8 & 3012 & 320 & 3887 & 24 \\
2018-12-22 & 8474.71165 & $-813$ & 8 & 3193 & 314 & 3901 & 24 \\
\multicolumn{8}{l}{\aql} \\
2014-02-15 & 6704.16035 & 219 & 4 & 1940 & 1336 & 1185 & 55 \\
2014-02-20 & 6709.16445 & 221 & 4 & 2199 & 1160 & 1177 & 52 \\
2019-03-19 & 8562.15905 & 143 & 3 & 928 & 2921 & 1194 & 55 \\
2019-06-09 & 8644.00385 & 143 & 3 & 2001 & 1554 & 1203 & 58 \\
\multicolumn{8}{l}{\leo} \\
2006-06-09 & 3895.80468 & $-259$ & 11 & 3382 & 280 & 2235 & 51 \\
2006-06-10 & 3896.81240 & $-268$ & 11 & 4247 & 182 & 2186 & 37 \\
2006-06-11 & 3897.80049 & $-257$ & 11 & 3705 & 180 & 2170 & 36 \\
2006-06-12 & 3898.77849 & $-287$ & 12 & 3905 & 165 & 2257 & 35 \\
\multicolumn{8}{l}{\lac} \\
2006-08-05 & 3953.07311 & $-432$ & 25 & 2394 & 556 & 1903 & 92 \\
2006-08-07 & 3955.06634 & 320 & 19 & 4182 & 473 & 3109 & 97 \\
2006-08-08 & 3956.06002 & $-353$ & 19 & 3948 & 361 & 3085 & 87 \\
2006-08-09 & 3957.05969 & $-393$ & 23 & 3452 & 460 & 1572 & 64 \\
\hline
\end{tabular}
\tablefoot{\bz\ is obtained from the first moment of $Z_1$.
The two determinations of $\mbsq^{1/2}$ come from the moment ratio of
Eq.~(\ref{eq:moments13}) and from the constrained deconvolution of
Sect.~\ref{sec:practical}. 
}
\end{table}

\begin{figure}[!t]
\figps{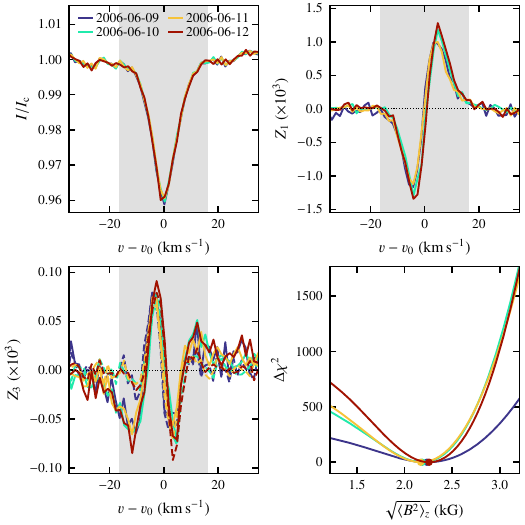}
\caption{Same as Fig.~\ref{fig:gamequ} for the four ESPaDOnS observations of \leo.}
\label{fig:leo}
\end{figure}

\begin{figure}[!t]
\figps{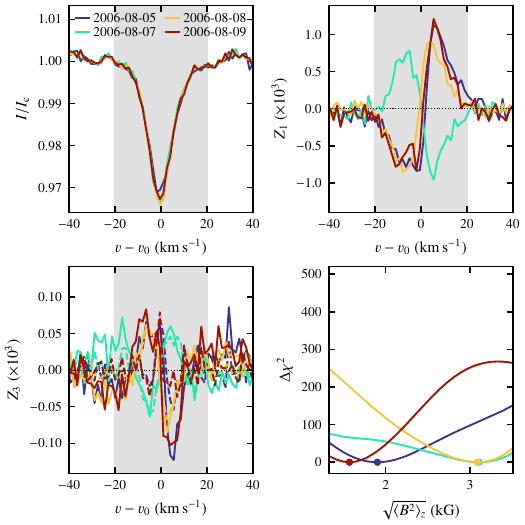}
\caption{Same as Fig.~\ref{fig:gamequ} for the four ESPaDOnS observations of \lac.}
\label{fig:lac}
\end{figure}

\subsection{Active M dwarfs \leo\ and \lac}
\label{sec:mdwarfs}

Both Ap stars carry a longitudinal field broadly consistent with their modulus in the framework of a smooth, global fossil magnetic field,
so that the second observable in principle largely follows the first. Active M dwarfs
are the opposite case, with a multi-kG field concentrated at small spatial scales against much weaker large-scale fields with
\bz\ of a few hundred gauss \citep[e.g.][]{kochukhov:2021}, and it is for them that Eq.~(\ref{eq:filling})
matters. Here we analyse representative sets of observations of two well-known active M dwarfs,
\leo\ (GJ\,388, dM3.5) and \lac\ (GJ\,873, dM4.0). For each of these targets we picked four
ESPaDOnS observations acquired in 2006, on nights covering a significant fraction of a rotation of 2.24 and
4.37~d \citep{morin:2008}, respectively, at $S/N$ = 200--450 per pixel.

Significant molecular absorption in these cool (\teff\ = 3300--3400~K) stars limits the adjustments 
that can be made to the mask. Molecular features
blanket these spectra below 8000\,\AA\ and contribute almost nothing to circular
polarisation, so all 3349 atomic lines between 5000 and 9850\,\AA\ that survive the
excluded windows and the 5\% depth cutoff are retained here, without any depth adjustment
and without the line depth threshold applied to the Ap stars. The broad
features are excluded from a tabulated list of their positions rather than by the
pseudo-continuum test of Sect.~\ref{sec:blendsetup}, whose running maximum never
reaches the true continuum in an M dwarf spectrum. 
Telluric bands are removed for calculation of \bz,
which requires $W_I$, and retained
for the second-order moment ratio and fit, which need no Stokes $I$ at all.

The second-order LSD Stokes $V$ profile $Z_3$ is detected in all eight M-dwarf spectra (Figs.~\ref{fig:leo} and
\ref{fig:lac}) at $\Delta\chi^2$ = 102 to 1135, with an amplitude 6 to 7\% of that of $Z_1$. 
This is an indication, obtained directly from the observed Stokes $V$ data, that polarised line formation in these two dwarfs does not occur in the weak-field regime despite their modest \bz.
Because $w^{(3)}/w^{(1)}$ grows as
$\lambda^2$, the fit takes its leverage from the reddest lines, which is also
where the atomic absorption of an M dwarf emerges from the molecular bands, so an error in the mask
varying smoothly with wavelength could in principle imitate the second component.
Two tests exclude this. The blue and the red halves of the mask detect $Z_3$
separately and agree to 19\%, and permuting $G^{(3)}_1$ among mask lines of
similar wavelength, which preserves the $\lambda^3$ scaling and the distribution
of $w^{(3)}$ exactly, removes the detection altogether, the fit returning
$\gamma$ consistent with zero in every trial. The detected second component is thus linked to the correct
Zeeman moment of each individual line.

The field moduli recovered from the constrained LSD inversion are 2212~G for \leo, with an rms scatter of 36~G over
four rotational phases, and 1572 to 3109~G for \lac, whose four values vary
with rotational phase much more than for \leo.
Both stand six to eight times above the longitudinal field of the same spectra, which is
the separation of field strength from filling factor of Eq.~(\ref{eq:filling})
obtained on a star instead of on a synthetic surface. They are also far closer to
the mean field moduli these stars are known to have from Zeeman broadening, about 2.6--3.3 and
3.7--4.2~kG \citep{kochukhov:2021}, than to the 0.2--0.3 and 0.5--0.6~kG mean strength of
their large-scale fields reconstructed from circular polarisation by ZDI \citep{morin:2008}.
This is the first model-independent observational evidence that the Stokes $V$ signal 
in active M dwarfs emerges from surface regions permeated
by kG-strength fields inferred from Zeeman broadening, a finding that invalidates the 
weak-field treatment on which the published ZDI maps rest \citep{kochukhov:2026}.
The residual difference goes in the direction of the truncation bias of
Sect.~\ref{sec:blenddet}, which these uncalibrated values do not correct for and
which brings \leo\ to 3.0~kG when the value measured at 2~kG (for the solar spectrum) is applied.
Two phases of \lac\ are a useful illustration of how the two observables derived from second-order 
Stokes $V$ LSD profile analysis may differ.
Between the two consecutive nights of 2006 August 7 and 8, \bz\ reverses from $+320$ to
$-353$~G, 25 times its error, whereas $\mbsq^{1/2}$ moves by 24~G against an
error of 130. The longitudinal field inverts but the field modulus estimator from the same Stokes $V$ signals is stable.

Two further points attach to these numbers. The moment ratio, listed beside
the fit in Table~\ref{tab:apstars}, exceeds it at every one of the eight phases,
by factors of 1.3 to 2.2. This is the ordering that Sect.~\ref{sec:blendfill}
predicts, the truncation of the expansion biasing the fit low while leaving the
third moment of $Z_3$ untouched, and the factors are those found on the synthetic
spectra at comparable field strengths.
The resulting moduli of 3.4--4.2~kG for
\leo\ and 2.4--4.2~kG for \lac\ are of the same order as the
Zeeman-broadening values quoted above, lying above them for the first star and
straddling them for the second.
The moment ratio is nevertheless not yet a converged measurement.
Its value has no plateau over the range of integration windows that the
profiles admit, growing by 55\% on \leo\ between the adopted 16\,\kms\ and the
22\,\kms\ out to which $Z_1$ remains significant.
Thus, the choice of window is a subjective judgement in this case, taken as 16\,\kms\ for \leo\ and 20\,\kms\ for \lac.
Additionally, the $Z_3$ profile of \leo\ exhibits a clear deviation from anti-symmetry and departs from scaled $Z_1''$ in the blue wing,
reaching there 60 to 80\% of its own peak and repeating in all four spectra.
This indicates that Eq.~(\ref{eq:tie}) does not hold in detail for this star. A
modulus varying over a disk that rotation resolves in velocity would produce such
a departure, which Sect.~\ref{sec:moments2} identifies as direct evidence of a
distribution of field strengths, but the neglected higher-order terms of the
expansion and the unmodelled polarisation of the lines outside the mask reach the
same wings, and these data do not unambiguously separate the three possibilities.

\end{appendix}

\end{document}